\documentclass[12pt,aps,prd,preprint]{revtex4} 
\usepackage{ulem}  
\usepackage{graphicx}

\usepackage{ifpdf,xspace,wrapfig,rotating,amssymb,amsmath,slashed,here,color,eso-pic}

\input{define_commands.input}
\begin{document}
\hspace{3.2in} \mbox{FERMILAB-PUB-10-028-E-PPD}


{\flushright \today}
{\flushright  Version 1.0}

\title{Jet Physics at the Tevatron}
\author{Anwar A Bhatti}
\address{The Rockefeller University, 1230 York Ave, New York NY 10065}
\author{Don Lincoln}
\address{Fermi National Accelerator Laboratory, P.O. Box 500, Batavia IL 60510}


\begin{abstract}
Jets have been used to verify the theory of quantum chromodynamics (QCD), measure the structure of the proton 
and to search for the physics beyond the Standard Model. In this article, we review the current status of jet 
physics at the Tevatron, a $\sqrt{s} = 1.96$ TeV $p\bar p$ collider
at the Fermi National Accelerator Laboratory. We report on recent measurements of the inclusive jet production cross section and the 
results of searches for physics beyond the Standard Model using jets. 
Dijet production measurements are also reported. 
\end{abstract}

\maketitle

\section{Introduction}
The theory of quantum chromodynamics (QCD)~\cite{theory_pQCD} is currently the best description of the fundamental strong force. 
This theory describes the color interaction between quarks as being mediated by gluons, which are the vector bosons of the strong force. 
It has been successfully tested in collisions between $e^+e^-$ \cite{lep_qcd}, $ep$ \cite{hera_qcd}, 
$pp$ \cite{isr_qcd}, and $p\overline{p}$ \cite{tev_qcd}.

One of the basic properties of QCD is that its coupling strength \alphas decreases with the energy of the interaction
and that, at sufficiently high energies, QCD calculations can be performed using perturbation theory in powers of \alphas~\cite{summary_2009_theory}.
Currently, these perturbartive QCD  (pQCD) calculations are available at a next-to-leading order (NLO) for many processes and, in some cases, at 
next-to-next-to-leading order (NNLO) approximation.
Leading order (LO) calculations, supplemented with parton shower calculations~\cite{parton_shower},
are used in several Monte Carlo event generators~\cite{pythia,herwig}. In addition, matrix element generators which match NLO 
calculations of rates for QCD processes with a parton shower Monte Carlo event generator are also available~\cite{MC-NLO},       but
only a limited number of processes have been implemented.
 
The pQCD calculations result in a small number of partons in the final state, while experimenters 
observe ``jets'' of particles.  These jets retain the kinematic properties (energy and momentum) of the parent partons (quarks or gluons).  
In order to facilitate comparison between data and calculation, jet finding algorithms have been devised 
that are insensitive to the difficult-to-calculate low energy phenomena that govern the transition from low-multiplicity partons 
to high-multiplicity particle final states.  There are several jet finding algorithms and the details of the measurements 
are sensitive to that choice.

While events in which jets are created are used for a detailed understanding of the strong force, 
it is also possible that such events could also reveal new physical phenomena, including quark substructure (compositeness), 
extra spatial dimensions and new particles  which decay into jets. 
Because of their high energy, jets can probe very small distances. At the Tevatron, the highest \pT jets
can probe distances down to ${\cal O}(10^{-17})$ cm.

In 2001, the Fermilab Tevatron $p\overline{p}$ collider commenced its Run II, with a collision energy of 1.96 TeV.  This energy is higher than 
the $1992-1996$ Run I energy of 1.8 TeV.
Even this relatively small increase in energy leads to a substantial increase in jet production with large transverse momentum, \pT,
by about a factor of three at \pT = 500 \GeVC.
The beam intensity is much higher than Run I due to the addition of the Main Injector and the Recycler Ring to the Fermilab accelerator complex.
In addition, both the CDF~\cite{overview_cdf} and \D0 detectors~\cite{overview_d0} were upgraded.
The results reported here utilize an order of magnitude higher integrated luminosity than reported previously~\cite{tev_qcd}.  

\section{Perturbative QCD\label{sec-pQCD}}
The theory of QCD describes the behavior of those particles (quarks $q$ and gluons $g$) that experience the strong force.  
It is broadly modeled on the theory of Quantum Electrodynamics (QED), which describes the interactions between electrically-charged particles.  
However, unlike the electrically-neutral photon of QED, the gluons, the force-mediating bosons of the strong interaction, 
carry the strong charge.  
This fact greatly increases the complexity in calculating the behavior of matter undergoing interactions 
via the strong force. 

The mathematical techniques required to make these calculations can be found in textbooks (e.g. \cite{Kieth}).  
Instead of giving an exhaustive description of those techniques here, we focus on those aspects of the calculations employed most frequently in the experimental analysis, 
thereby clarifying the phenomena experimentalists investigate.

\begin{figure}[htbf]
\centering
\includegraphics[width=120mm]{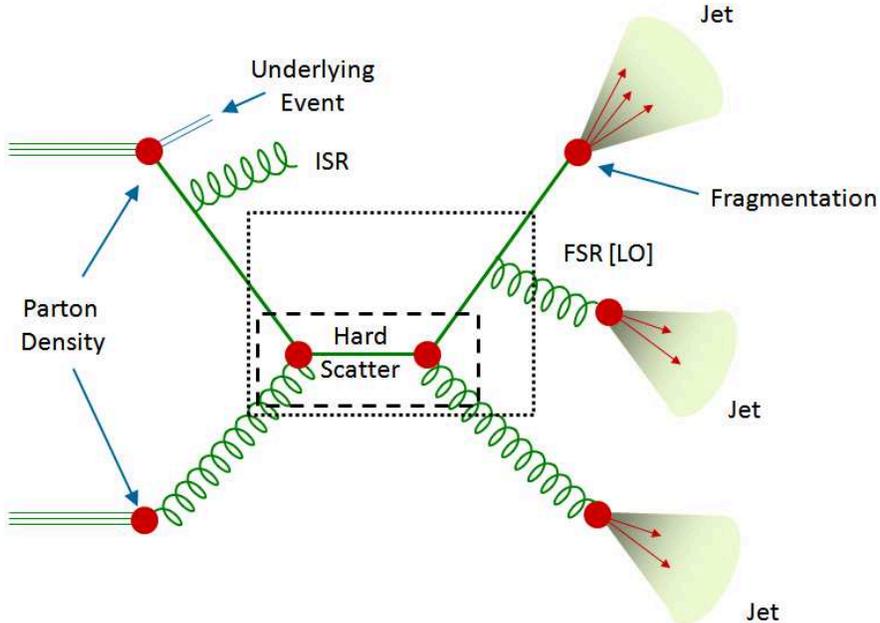}
\caption{Stylized hadron-hadron collision, with relevant features labeled. Note that a LO calculation of the hard scatter (dashed line) 
will assign a jet to final state radiation that would be included in the hard scatter calculation by a NLO calculation (dotted line).}
\label{Event_cartoon}
\end{figure}

\begin{figure}[htbf]
\centering
\includegraphics[width=120mm]{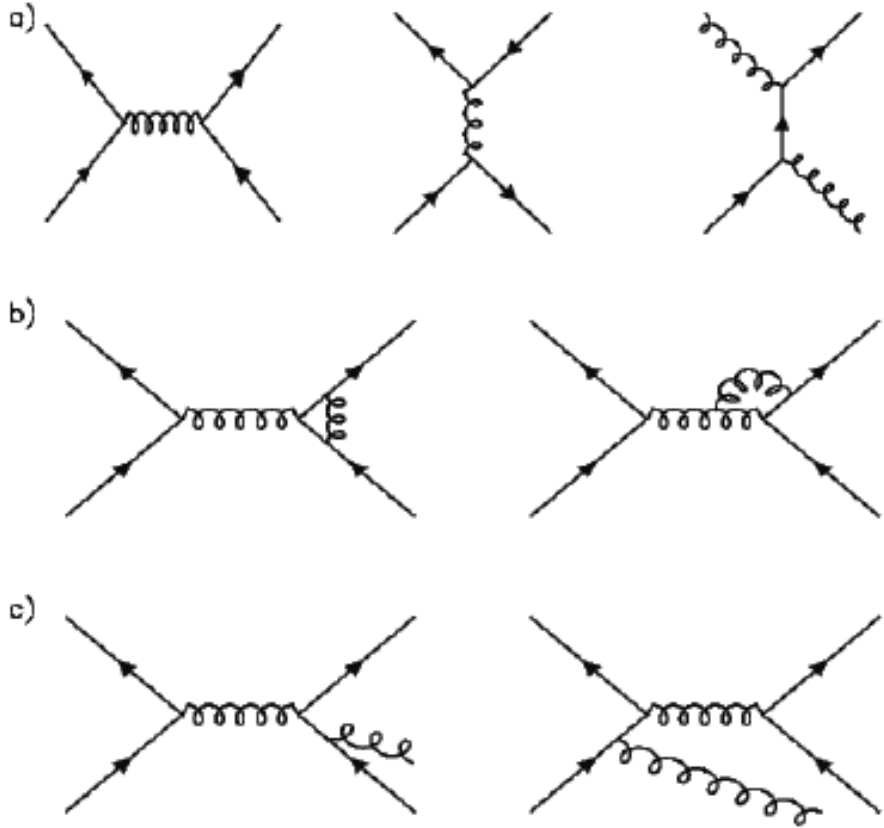}
\caption{(a) Leading order Feynman diagrams. (b) Next to leading order loop diagrams.  (c) Next to leading order tree diagrams.} 
\label{fig:feynman_LO_NLO}
\end{figure}


At high energies, the strong interactions between two hadrons can be factorized into three components: (a) the probability of finding the partons
in the hadrons, (b) the interaction between quasi-free partons, and (c) fragmentation and hadronization of the final state partons. 
The process is schematically shown in Figure \ref{Event_cartoon}. The cross section of the hadron-hadron scattering with 4-momenta $P_1$ and $P_2$ 
can be written as~\cite{Kieth} :
\begin{equation}
\sigma(P_1,P_2) = \sum_{i,j}\!\!\int\!dx_1\int\!\!\!dx_2\;f_i(x_1,\muF^2) f_j(x_2,\muF^2) \hat{\sigma}_{ij}(p_1,p_2,\alpha_s(\mu_R^2),Q^2/\muF^2,Q^2/\muR^2) 
\label{eqn-hard_scatter}
\end{equation}
The hard interaction between partons $i,j$ is given by $\hat{\sigma}_{ij}$ where $p_1=x_1 P_1$ and  $p_2=x_2 P_2$ are
4-momenta of the two partons. It is independent of the incoming hadrons' structure and can be calculated using pQCD.
The function $f_i(x,\muF^2)$ is the probability of finding a parton $i$ with momentum fraction $x$ at
the scale $\muF$, and is called the parton distribution function PDF. 
The sum $i,j$ is over the partons in the two respective hadrons.
$\alpha_s$ is the strong coupling constant.
$\mu_R$ is the renormalization scale, which is used to remove non-physical infinities inherent in fixed-order calculations. 
$Q$ is the characteristic scale of the interaction and is related to some physical scale in the interaction, such as the \pT of the leading jet.

This factorization of hadron-hadron interactions into a short distance interaction (hard interaction) and a large distance interaction (PDFs)
is done at an arbitrary energy scale \muF. A parton emitted below the scale \muF is considered to be part of the hadron structure and thus is described by the PDFs.
The hadron structure is measured by many different experiments, especially lepton-hadron scattering experiments.
The PDFs are determined by fitting these data and parametrized at a scale $\mu_0$. 
The QCD evolution equations~\cite{parton_shower}, currently available at NNLO in perturbation theory, are used 
to evolve these PDFs to any different scale $\mu$.
The cross section is a convolution of the PDFs and the parton level cross section and, to be consistent, both must
be calculated to the same order in perturbation theory.
A complete calculation, including all orders in the perturbation series, must be independent of \muF, but a fixed order calculation depends on this choice.
To evaluate the sensitivity to \muF, it is usually varied up and 
down by a factor of two. Scale dependencies are expected to decline with the addition of higher order terms in the calculation.

In Figure \ref{fig:feynman_LO_NLO}, for jet production, examples of leading order (LO) and
next-to-leader order (NLO) Feynman diagrams are shown.  
At NLO, the jet cross section receives contributions from virtual corrections to the two parton final state, 
and from real corrections from the three parton final state. Both contributions are divergent, but the sum is finite.
One sees that a NLO pQCD calculation can describe up to three jets in the final state.
For inclusive jet production, many different implementations of NLO pQCD calculations are available~\cite{EKS,JETRAD,NLOJet++,FastNLO}.
All these programs use  Monte Carlo integration techniques to calculate the real and virtual contribution to the cross sections. The EKS program~\cite{EKS}
calculates the cross sections for user-predefined cuts on transverse momentum and rapidity ranges. 
JETRAD~\cite{JETRAD} and NLOJet++\cite{NLOJet++} generate parton events with weights (both positive and negative) and thus full event kinematics 
are available to the user for jet clustering, detector acceptance calculations and study of any other distribution. 
These programs require a huge amount of CPU to reach the desired precision.  A large fraction of this CPU is consumed in evaluating the
PDFs. In the FastNLO program~\cite{FastNLO}, the convolution in Equation~\ref{eqn-hard_scatter} is modified to 
a product of a perturbatively calculable piece $\tilde\sigma_{n,i,k,l,m}(\mu)$, calculated using NLOJet++, 
the strong coupling constant $\alphas^n(\mu^{(m)})$, and a function $F_i(x_a^{(k)},x_b^{(l)},\mu^{(m)})$ which depends on only the PDFs 
and the factorization scale $\mu$. The time consuming piece $\tilde\sigma$ is calculated only once. 
The function $F_i$ is calculated on a grid of 
different values of $x_1$, $x_2,$ and $\muF$ only once and interpolated between those points while evaluating the full cross section.
This procedures significantly speeds up the calculations.

Near threshold, the phase space for the emission of real gluons is 
limited and large logarithmic corrections to the above cross section calculation may arise from the incomplete cancellation of infrared divergences
against the virtual gluon emission contributions. For jet production, these corrections are expected to
contribute at very high $x$, where parton distributions are falling very steeply. 
For the inclusive jet cross section, these threshold corrections have been calculated to NNLO at next-to-leading-logarithmic accuracy~\cite{QCD-2-loop}
and are found to be small. However, the corrected cross section shows a substantial reduction of 
the scale \muF,\,\muR dependence.

The partons radiate when they pass through a color field. 
In this type of radiation, two forms dominate, collinear and soft radiation. Collinear radiation is in the direction of
parent parton, while soft radiation is just low energy emission. These soft and
collinear radiation can be calculated in a leading-logarithmic approximation to all orders 
and this algorithm is a crucial component of event generators.
In this formulism, a gluon radiates another gluon or
converts into a $q\bar q$ pair according to DGLAP (Dokshitzer-Gribov-Lipatov-Altarelli-Parisi) splitting 
functions~\cite{parton_shower} which depend on \alphas and a variable $z$ which describes the energy sharing between two
daughter partons.
Similarly, a quark or an anti-quark can radiate a gluon which radiates further. This sequential
radiation results in a shower of partons. The radiation process is continued until the parton virtual mass $t$ is smaller than a mass scale $t_0$.
This simple procedure is augmented by angular ordering, i.e. each subsequent emission is required to have a smaller angle, to simulate color coherence effects. 
Color coherence leads to suppression of soft gluon radiation in certain regions of phase space. In the final state showers, the radiation is limited to
a cone defined by color flow lines and the emission angle at each branch point is smaller than the previous emission angle.
Partons
with a virtual mass $t\le t_0\sim 1$ \GeVCsqr are combined into hadrons using a phenomenological hadronization model. 
The hadronization models have been tuned to reproduce the jet structure observed at $e^{+}e^{-}$ colliders.
The showering process described above is used for radiation from outgoing partons and called final state radiation (FSR).
For radiation from incoming partons, initial state radiation (ISR), the event generators use so-called backward evolution. 
First the momentum fractions $x_1,x_2$ of partons 
participating in the hard interaction are determined. Then, the parton shower that preceded the hard interaction 
is subsequently reconstructed,
evolving partons from the hard interaction scale $Q$ backward in time towards smaller $Q$ where the PDF $f$ is evaluated.
The color coherence in initial state radiation is slightly more complicated but still follows the same angular ordering. 
Interference between initial state and final state radiation is implemented in \herwig
but not in \pythia.
The independent variable, $t$, which is how the evolution of $\alpha_s(t)$ is parametrized, is not unique. 
In the \pythia showering algorithm, the squared mass $m^2$ of the branching parton is used as the evolution variable.
\herwig uses $t = m^2/(2z(1-z))$ where $z=E_b/E_a$ is ratio of daughter parton energy ($E_b$) to the parent parton energy ($E_a$). 
In recent versions of \pythia, the option of $t=\pT$ of the branching parton is also available
as the evolution variable.

Occasionally, the radiated parton is at a sufficiently large angle to the parent parton and carries enough energy that it leads to an identifiable jet.
Because these jets are typically of lower \pT, it does not dominate the event kinematics.
However, this lower $p_T$ radiation becomes important in studies of jet multiplicity. This part of ISR/FSR can be considered as a  part
of the hard scatter $\widehat\sigma_{ij}$ or treated independently.
For instance, we see in Figure~\ref{Event_cartoon} two boxes surrounding the hard scatter.  
The dashed box surrounds the leading order scatter, while the dotted box surrounds a next-to-leading order diagram.
At NLO pQCD, one parton emitted from either incoming or outgoing parton is part of the short distance hard cross section $\widehat{\sigma}_{ij}$.
In LO event generators, these parton emissions are treated quasi-independently of the hard scatter and are part of
leading-log showering process.

The partons from the incoming hadrons which do not participate in the hardest scatter in an event also interact, but these interactions 
are normally soft. The particles produced in these multi-parton interactions are, on average, isotropically distributed 
in the allowed $y-\phi$~\cite{kinematics} space and can overlap with the jets produced in the hard interactions.
Being soft, multi-parton interactions are in the non-perturbative regime and thus are
implemented in event generators using phenomenological models. 
The parameters of these multi-parton interaction models have been tuned to reproduce the transverse energy and multiplicity
distributions of the particles observed far away from the hard jets in collider data~\cite{Rick-TuneA,QW-tune}.
The beam remnants, the partons which do not participate  either in hard or multi-parton interactions, go along the beam direction.
However, they do carry color and to become color-singlets, they must exchange (soft) partons with the rest of the event. 
The particles produced in multi-parton interactions and from the hadronization of beam remnants
collectively constitute the underlying event. 

As described above, pQCD predictions for jet production are available at NLO at the parton level only. These predictions can not, in principle, 
be compared directly with the data which is available at the particle level, 
because these parton level calculations do not include hadronization effects and the contribution from the underlying event.
On the other hand, event generators include hadronization, underlying event energy, and ISR/FSR to all orders in the leading-log approximation, but the
hard interaction is calculated at LO only. 
Thus to compare data with QCD predictions, a hybrid scheme is generally used.
The parton level NLO calculations are correcting for the underlying event and hadronization effects before
they are compared with data.
These corrections are determined using Monte Carlo event generators 
by comparing the jets obtained by clustering the pQCD partons with the jets obtained by
clustering the partons after the showering process. The parton jets are obtained from the Monte Carlo events
in which the underlying event simulation (multiple parton scattering) has been turned off.

Over the last decade, there has been a lot of progress in simulating high jet multiplicity events using tree-level matrix elements.
In \alpgen~\cite{alpgen}, events with the exclusive parton multiplicities $n=2,3,4$, and $5$
are generated using matrix elements from pQCD at the tree level. To include the effect of soft and collinear emission to all orders (albeit in the leading-log approximation),
these events are passed through a showering program e.g. in \pythia. The phase spaces of the matrix elements and the parton showering program overlap.
In particular, showering programs occasionally generate hard partons which can lead to a state which has already been generated by the matrix element.
To avoid this double counting, a matching criteria is used. For example, in \alpgen it is required that the number of jets produced by clustering the partons 
produced by the matrix element is the same as those produced after the showering for $n\le 4$. The events which do not satisfy this condition are rejected. 
For the $n=5$ parton state, the showering algorithm is allowed to produce a higher jet multiplicity state. The spectra from each different multiplicity are
combined to form the full spectrum.
In \sherpa~\cite{sherpa}, a different matching procedure~\cite{ckkw} is used where parton showers above a cut off \kT-like
measure (c.f. Equation ~\ref{eqn-kt}) are vetoed. Both \alpgen and Sherpa have been extensively tested at the Tevatron in $W/Z$+jet production,
but these studies are not discussed here due to space constraints~\cite{EllisReview}.

While lepton-nucleon deep inelastic scattering (DIS) experiments are able to precisely measure the quark content of 
the proton, this precision is not achievable for the gluon, especially at high $x$. At low $x$, the gluon distribution can be determined precisely using 
QCD scaling violation in DIS data. Studying the high-$x$ gluon distribution functions requires data from hadron-hadron 
scattering. The effect of including Tevatron jet data in global fits to determine PDFs is described in Section~\ref{sec-pdfs}. 

\section{Jet Clustering Algorithms}
Because a parton carries the strong charge, it is not directly observable. It showers into many partons which combine together to 
form a large number of particles which travel in roughly the same direction as the initial parton. 
The kinematic properties of the initial parton can be inferred either from the shower of partons or from the jet of collimated particles. 
For this inference, these particles or partons must be clustered into a {\it jet} by an algorithm. 
In pQCD, at NLO and higher orders, a jet algorithm is needed to define physics observables which are well-defined i.e. they are 
soft/collinear safe. Jet algorithms are run on a few partons generated in pQCD calculations to construct such variables. 
Experimentally, the final state particles are observed as tracks in the tracker systems or as towers of energy in the calorimeter. 
These tracks or towers  must also combined into a jet so that they can be compared to the parton produced in the hard interaction.
In the following, we will collectively call the (a) partons in a pQCD calculation, (b) partons or (c) particles  produced in Monte Carlo event
generators, or (d) towers or (e) tracks, or (f) reconstructed particles observed in a detector as the {\it objects} which are
input for a clustering algorithm.

For a valid comparison between observations and theoretical predictions, the clustering algorithm 
must satisfy some basic criteria~\cite{Run2JetPhysics,LesHouche}. 
The algorithm must be safe against soft (infra-red IR) and collinear radiation, invariant to boosts along the beam direction, and should be insensitive 
to the non-perturbative hadronization effects. In an algorithm which is not safe against soft/collinear radiation, 
the virtual and real contributions in pQCD calculations do not cancel completely and thus the predicted cross sections 
are ill-defined. One should be able to run the same algorithm on the detector calorimeter towers or tracks, particles or the multi-parton state 
from event generators and partons in fixed order pQCD calculations and get sensible results.
Experimentally, jet clustering algorithms
should be insensitive to the energy from additional hadron-hadron collisions in the same
bunch crossing which overlaps the energy from hard interactions, and should not consume too much computer resources such as CPU. 
Finally, the algorithm must be completely specified to avoid different interpretations.

Commonly used jet-finding algorithms can be divided into two categories: (a) cone clustering and 
(b) pair-wise recombination algorithms. With a few exceptions, only cone clustering algorithms 
have been used at hadron colliders. The cone clustering algorithms used prior to the Tevatron Run II were 
not IR/collinear safe~\cite{Run2JetPhysics,ThreeJet},
and it was proposed to add an additional seed at the midpoint of stable cones. This made the new algorithm 
IR/collinear safe to NLO for the inclusive jet cross section
measurement. For other physics observable, it is either safe at  LO only or unsafe at all orders~\cite{LesHouche}.
Various issues related to jet reconstruction are extensively discussed in a recent review~\cite{Jetography}.

\subsection{\bf Cone Clustering Algorithm}
For jet studies in Run II, both the CDF and \D0 collaborations are using the Midpoint algorithm, as laid out by the 
QCD Workshop recommendations~\cite{Run2JetPhysics}, but the two implementations
differ in some details. Below we describe the implementation of this algorithm by the CDF collaboration.

The clustering process starts by making a list of all objects to be clustered. 
In simulated events, all the particles or partons are included without a \pT threshold. However, in data,
the calorimeter towers are required to have $\pT\ge 100$ \MeVC to minimize the effect of noise. From this list, a second list of
seed objects is made with the requirement that the \pT of the objects exceeds a fixed threshold of 1.0 \GeVC. At each seed location,
the 4-momentum of the cluster is determined by summing the 4-momenta of all the objects within a distance 
$R=\sqrt{(y-y_c)^2+(\phi-\phi_c)^2}$ from the seed ($y_c,\phi_c$). The 4-momenta are summed using the E-scheme \cite{Run2JetPhysics},
\begin{eqnarray}
(E,p_x,p_y,p_z) = \sum_{i}(E,p_{x},p_{y},p_{z})_i \\
 p_T = \sqrt{p^2_x+p^2_y}\hspace{5ex}
y_c =  \frac{1}{2} \ln \left(\frac{E+p_z}{E-p_z}\right)\hspace{5ex}
\phi_c = \tan^{-1}(p_y/p_x).\label{eqn-pt-y-phi}
\end{eqnarray}
This scheme is different from the Snowmass scheme \cite{snowmass_algorithm} used in Run I, where the clustering centroid was 
defined as the $E_T$-weighted average
of $\eta$ and $\phi$. 
Using the center of the cluster as a new seed location, the process is iterated until the center of the circle $(y_c, \phi_c)$ 
coincides with the position of cluster 4-momentum.

After all the stable cones have been identified beginning with real seeds, there is an additional search for stable cones 
using as seed locations the midpoints between the initial set of stable cones.  The cone finding algorithm allows that the 
same object may be part of many cones. The shared objects are uniquely assigned to a single cone
using a split-merge algorithm as specified in~\cite{Run2JetPhysics}. 
If two stable cones share objects, the shared transverse energy is 
compared to the transverse energy of the lower \pT cone.  
If the ratio of the shared transveres energy to the transverse  
energy of the lower \pT cone is higher 
than the energy fraction $f_{merge}$,
the two cones are merged. Otherwise, based on proximity, the shared objects are assigned to the nearest cone.
The two collaborations use different values of $f_{merge}$: CDF (\D0) uses $f_{merge}=$ 0.75 (0.50). 
This split-merge procedure may lead to jets which are not circular in $y-\phi$ space.
After all the objects above threshold have been uniquely assigned to a stable cone, the jet kinematics are determined using the 
same E-scheme.

A cone clustering algorithm can be made infrared safe to all orders if a stable cone is evaluated at each point in $\y-\phi$ space. 
Such an algorithm is very CPU-intensive even when the number of
particles is modest and thus is not practical beyond some parton level pQCD calculations.
Recently, a new seedless cone clustering algorithm has been proposed which is infra-red and collinear safe to all orders
in perturbation theory. 
The Seedless Infrared Safe Cone {\sc siscone} algorithm~\cite{SISCone} uses the fact that a circle enclosing a set of particles can be moved around
such that two of the particles lie on its circumference. 
Consequently, all stable circles can be reconstructed by
considering all possible pairs of particles. After determining all the stable circles h, the
algorithm merges and splits the stable circles to uniquely assign the particles to a single circle.  This algorithm is fast and has been
used at the Tevatron for comparison purposes only.

\subsection{Pairwise Clustering Algorithm}
The cone algorithm combines all the objects within a distance $R$ from the seed. 
In contrast, the recombination
algorithms combine pairs of objects based on some measure $d_{ij}$ and is an attempt to ``undo'' the showering of partons.
The \kT algorithm~\cite{theory_kt_1,theory_kt_2} starts with a list of proto-jets given by 4-momentum $(E,p_x,p_y,p_z)$. 
All the objects which are to be clustered are considered as proto-jets.
The transverse momentum \pT, rapidity \y, and azimuthal angle $\phi$ of a proto-jet are calculated using Equation~\ref{eqn-pt-y-phi}.

For each proto-jet $i$ and the pair $(i,j, \; i\neq j)$, $d_i$ and $d_{ij}$ are defined as
\begin{eqnarray}
d_i = p_{T,i}^2 \hspace{8ex} d_{ij} = \min(p^{2p}_{T,i}, p^{2p}_{T,j})\;\frac{(y_i-y_j)^2+(\phi_i-\phi_j)^2}{D^2}
\label{eqn-kt}
\end{eqnarray}
where $D$ is the parameter which controls the size of the jet. For the \kT algorithm, the parameter $p=1$.
The algorithm determines the minimum $d_{min}$ of
the $d_i$ and all the $d_{ij}$. If $d_{min}=d_i$, the proto-jet is not mergable and is promoted to a jet. Otherwise, the
proto-jets $i,j$ are merged into a single proto-jet with the 4-momentum $(E_{ij}, \vec{p}_{ij})=(E_i+E_j,{\vec{p}_i+\vec{p}_j})$.
The process is repeated until no proto-jets are left.

The \kT algorithm has been extensively used at $e^+e^{-}$ and $ep$ colliders. At hadron colliders, the environment is
more challenging. The energy from
multi-parton interactions and beam remnants and pile-up can contribute to the jets and must be taken into account.
The large particle multiplicity observed in hadron-hadron collisions requires substantial CPU resources to process an event.
Thus the use of the \kT algorithm has been limited at hadron colliders. 
The \D0 collaboration measured the inclusive jet cross section in Run I \cite{D0-ktRunI}.
In the Tevatron Run II, the \kT algorithm has  only been used by the CDF Collaboration to measure the inclusive jet cross section (described in Section~\ref{sec-kt-incjets}).

Recently, two more recombination algorithms using $p\,=\,0$ (Cambridge-Aachen) \cite{Cambrige-Aachan} and $p=-1$ (anti-\kT)~\cite{AntiKt} in Equation~\ref{eqn-kt} 
have been proposed. The Cambridge-Aachen algorithm combines particles based only on their relative distance.
The anti-\kT algorithm combines the highest \pT objects in the events first. This leads to circular jets, which have well-defined area
like the cone jets. Thus far, these algorithms have not been used at the Tevatron.

\section{Jet Energy Scale Determination}

At the Tevatron, jets are generally measured using a calorimeter, which is sensitive to both charged and neutral
particles. Both CDF and \D0 utilize sampling calorimeters, which measure
only a small fraction of the energy of the particles. This observed energy is multiplied by a calibration constant so that it is equal to
the sum of the energies of the incident particles.
The calorimeter response is different for
hadrons, photons, electron and muons. For hadrons, the response depends on the momentum and the flavor of the particles, whereas
for photons and electrons it is almost momentum-independent. Muons normally deposit a little energy ($\sim 1$ GeV) in the calorimeter, which is almost
independent of the muon momentum. Neutrinos escape without interacting and lead to an imbalance in the measured \pT in the event.
The observed jet energy must be corrected for the calorimeter response
and other detector effects. The two collaborations employ different techniques to determine these jet energy scale
corrections. The CDF collaboration's technique~\cite{CDF-JES} depends on an accurate modeling of the calorimeter response to single particles and a knowledge
of the \pT spectrum of the particles in a jet,
whereas the \D0 technique is data-driven and utilizes the fact that in photon-jet events $p_{T,Jet}=p_{T,\gamma}$~\cite{D0-JES}. 
These techniques are used to calibrate the central region of the calorimeter where the tracking system is available to measure the charged particle
momentum and also the calorimeter response is uniform.

This approach was applied in the optimum calorimeter region for both collaborations.  The calorimeter response was extended to other regions ($0.1<|\eta|$, $|\eta|>0.7$ for CDF and $|\eta|>0.5$ for \D0), by using dijet balancing to scale the
jet energy response in the other regions to the one in the optimum region. The energy from additional $p\bar p$ interactions
in the same bunch crossing is subtracted, based on the number of reconstructed primary vertices in an event. For cone jets, this correction is
determined from minimum bias events by summing the energy in towers in a cone of radius $R$ placed randomly in the calorimeter. 
The procedure  for \kT jets is described in Section~\ref{sec-kt-incjets}.

\paragraph{\bf Photon-Jet balancing:} In this technique, the jet energy is determined by scaling the measured jet \pT to the 
photon's \pT in photon-jet events. The photon energy is measured by the electromagnetic (EM) calorimeter which is linear and has very
good energy resolution. In the approximation of
$2\rightarrow 2$ scattering, the jet transverse energy \pT is equal to the photon \pT. The real situation is
a little more complicated due to presence of initial state radiation ISR, the energy not 
clustered in the jet, and contributions to the clustered jet from the multiple parton interactions. 
To be insensitive to these effects (especially ISR), \D0 evaluates the missing \pT (\Met) projection fraction along the photon direction using:
\begin{eqnarray}
R_{had} =  1+ \frac{\vec{\Met}\cdot \vec{p}_{T,\gamma}}{p^2_{T,\gamma}}
\label{eqn-MPF}
\end{eqnarray}
The hadronic recoil correction factor, $R_{had}$, is the scale factor to the entire recoil system. By
requiring that the jet is back-to-back  with the photon and, in the absence of
any additional jet(s) in the event, $R_{had}$ is almost equal to the jet
response. The derived response is expressed in the jet energy \Eprime determined from
the \pT of the photon and the position of the balancing jet using 
$E^\prime \equiv \pT^{\gamma}/\sin\theta^{\rm jet}$
as both $\pT^{\gamma}$ and the direction of the jet
are accurately measured and thus \Eprime provides a  better estimate of the jet energy than the direct jet energy measurement by the calorimeter. 
It is preferred over jet \pT, as the calorimeter response depends on the energy of the incident particles and thus
parametrization of calorimeter response in \Eprime is more natural.

The EM calorimeter is calibrated using the electrons from $Z$ boson decays such that the reconstructed $Z$ boson mass 
is equal to the world average~\cite{pdg}. The EM calorimeter response to electrons and photons is similar, but not the same, as photons start
their shower later than electrons. This difference is small and is evaluated using simulated events at $\pT = 100$ \GeVC. The estimated
uncertainty on the photon energy scale is 0.5\% at low $E^\prime$ and 0.8\% at high  $E^\prime$.
 Using this procedure, the \D0 collaboration
has achieved a 1\% accuracy on the jet energy scale in photon-jet events. The current statistics of 
the $\gamma$+jets sample limit the direct measurement of
the jet energy corrections in the central region to $E^\prime<350$ GeV. The response is extrapolated to
higher energies using Monte Carlo, which has been tuned to the data. 
The correction to a single jet with a given algorithm and size
is deduced from $R_{had}$ using simulated events.

The calorimeter response to jets depends on their flavor, as the particle spectrum and multiplicity for quark-initiated and gluon-initiated jets are
different. The jet energy scale corrections determined from $\gamma$+jet events is valid only for the flavor composition of 
$\gamma$+jet events. Event topologies with different flavor composition will have different jet energy scale corrections.
Indeed, D{\O} tuned the single-pion response in their detector simulator to data and used \pythia to generate photon-quark, photon-gluon and
dijet events. They found that the gluon jet response was 8(2)\% lower for jets with 20(500) GeV of energy.  
In QCD jet production, the fraction of gluon-initiated jets changes with jet \pT and the corrections were adjusted to account for this variation in flavor composition.
With these additional corrections, the uncertainty on the jet energy scale is reduced to an unprecedentedly-small value.

\paragraph{\bf Jet Corrections using Single Particle Response:}
Another approach to determine the jet energy correction is based on a knowledge of the calorimeter response to
each particle that makes up a jet.  The CDF collaboration measured the calorimeter response to charged hadrons
and electrons using both $p\bar{p}$ collider and test beam data. The calorimeter simulation was tuned to reproduce
the measured response. 
The calorimeter response to a jet was determined by a convolution of the single particle response
with the type and momentum distribution of particles  constituting a jet as given by a fragmentation model. CDF used QCD dijet events with the 
\pythia fragmentation model to measure the default jet corrections.
The \pythia fragmentation model  agreed well with the particle \pT and multiplicity distributions in a jet measured 
in $p\bar p$ data. The {\sc herwig} event generator was used to crosscheck the \pythia fragmentation
functions and the results determined using two generators were found to agree well. In this procedure, the difference in calorimeter
response to gluon-initiated and quark-initiated jets is automatically included.
Although this procedure requires a detailed knowledge of
the calorimeter response and a well-tuned simulation, it has the advantages that the correction can be easily determined
for any event topology over the entire kinematic range, and real and simulated data have the same 
corrections and thus can be treated on an equal footing.

\section{Inclusive Jet Cross Section\label{sec-incjets}}
The inclusive jet cross section measurement  \cite{UA1-IncJets, UA2-IncJets, CDF-IncJet-Run0, CDF-IncJet-RunIa,CDF-IncJet-RunIb, D0-IncJet-RunI} 
has been used to test QCD and to search for physics beyond the 
Standard Model by searching for an excess of events 
at large \pT. During Run I at $\sqrt{s}=1.8$ TeV,
the search was limited by both theoretical and experimental systematic uncertainties even with an integrated luminosity of 100 \pbinv. The uncertainty on the jet energy scale dominated the experimental uncertainty. 
NLO pQCD calculations~\cite{EKS,JETRAD,NLOJet++} significantly reduced the dependence on the factorization and renormalization scales and 
the remaining dependence is $\sim 10\%$ and almost independent of jet \pT for \pT $>100$ \GeVC. 
The jet cross section is not very sensitive to non-perturbative hadronization effects. The underlying event contributes approximately 2 \GeVC of \pT to
a jet and thus is significant only at low \pT.
The main theoretical uncertainty arises from uncertainty in the parton distribution functions, especially for large $x$ gluons.
Unfortunately, inclusive jet production is the only process in which the high $x$ gluon distribution can be directly measured.
The other possible process, photon-jet production, has a limited reach in $x$ and the associated theoretical
uncertainties are large. The gluon distributions are also measured from lepton-proton scattering data through QCD scaling
violations, but these measurements are also limited to low $x$ values.
Because of these limitations,
the Run II inclusive jet cross section has been primarily used to constrain
the gluon content of the proton. 
The data at high \y are particularly useful as it probes high $x$ at lower $Q$ values 
where the contribution of physics beyond the Standard Model, 
if any, is negligible.

Due to the higher center of mass energy and much larger integrated luminosity, Run II jet measurements extend the jet spectrum to
higher jet \pT compared to the Run I measurements, by approximately 200 \GeVC. 
Both collaborations implemented an improved jet clustering algorithm. The jet clustering algorithms used in Run II
are IR/collinear safe at least to the order of the available pQCD calculations. An accurate determination of the calorimeter response from
the $p\bar p$ data and also a refinement in the techniques to determine the jet energy scale have lead to reduced uncertainty compared to
Run I.  In previous inclusive measurements, the hadronization effects were ignored as they were much smaller than 
both the experimental and theoretical uncertainties. In Run II, both collaborations have evaluated the effect of
hadronization and corrected the parton level pQCD calculation. In addition, the pQCD calculations are corrected for
the energy from the underlying event, determined using tuned event generators.  
In contrast, in Run I energy from the underlying event was removed from jets in data.
 
The CDF collaboration measured the inclusive jet cross section using the cone clustering~\cite{CDF-IncJet-Cone} 
with cone size $R=0.7$ and the \kT clustering
algorithm with $D=0.5,\,0.7\,$ and 1.0~\cite{CDF-IncJet-kt}. 
The \D0 collaboration has recently published the inclusive jet cross section~\cite{D0-IncJets} using
cone clustering algorithm with cone size $R=0.7$.  These three measurements are described below.

\subsection{Measurement using Cone Clustering Algorithm}
\subsubsection{\D0 Collaboration\label{sec-incjet-d0}}
\begin{figure}[htbp]
\begin{center}
\includegraphics[width=0.45\hsize,clip]{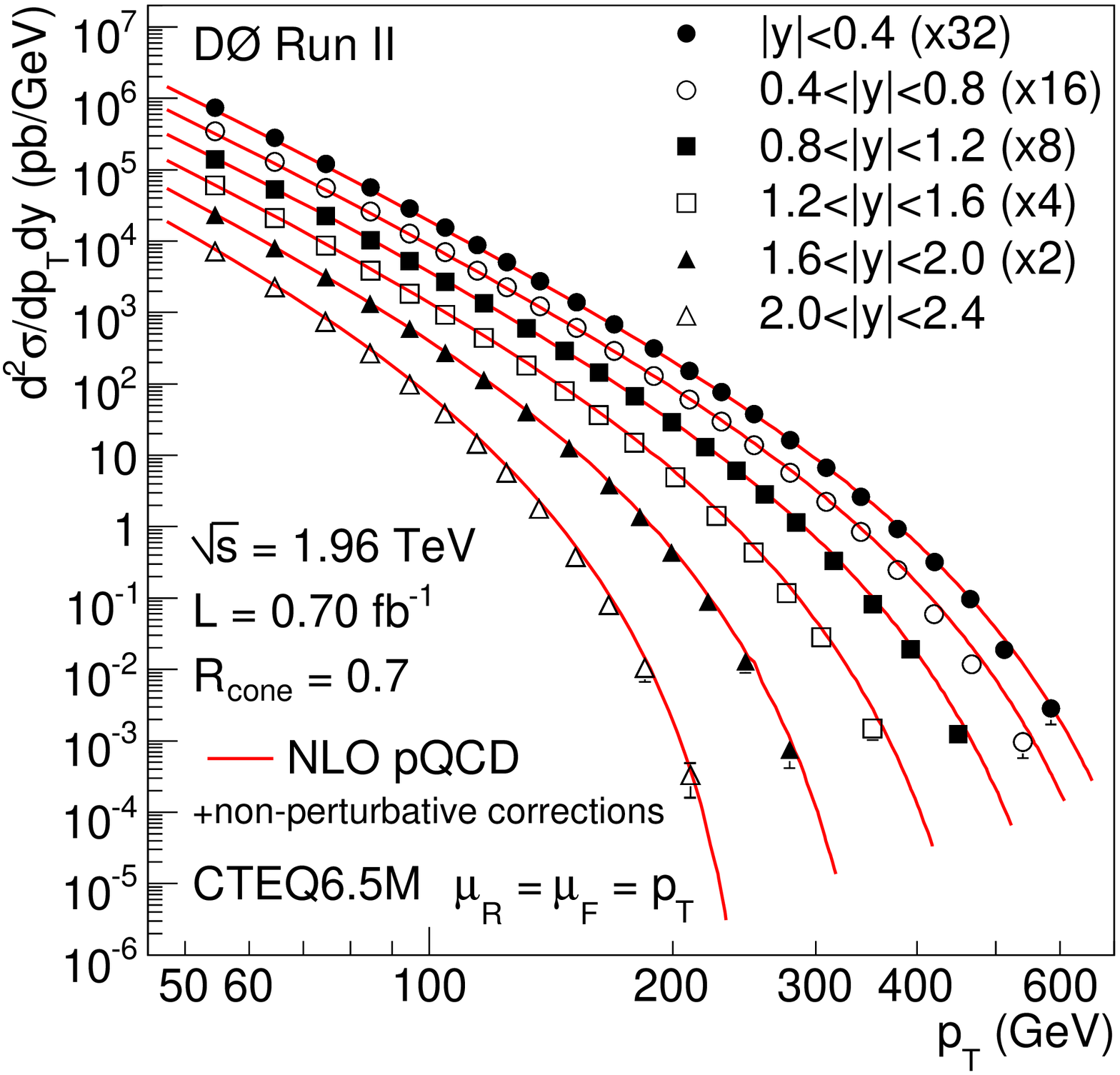} \\
\includegraphics[width=0.45\hsize,clip]{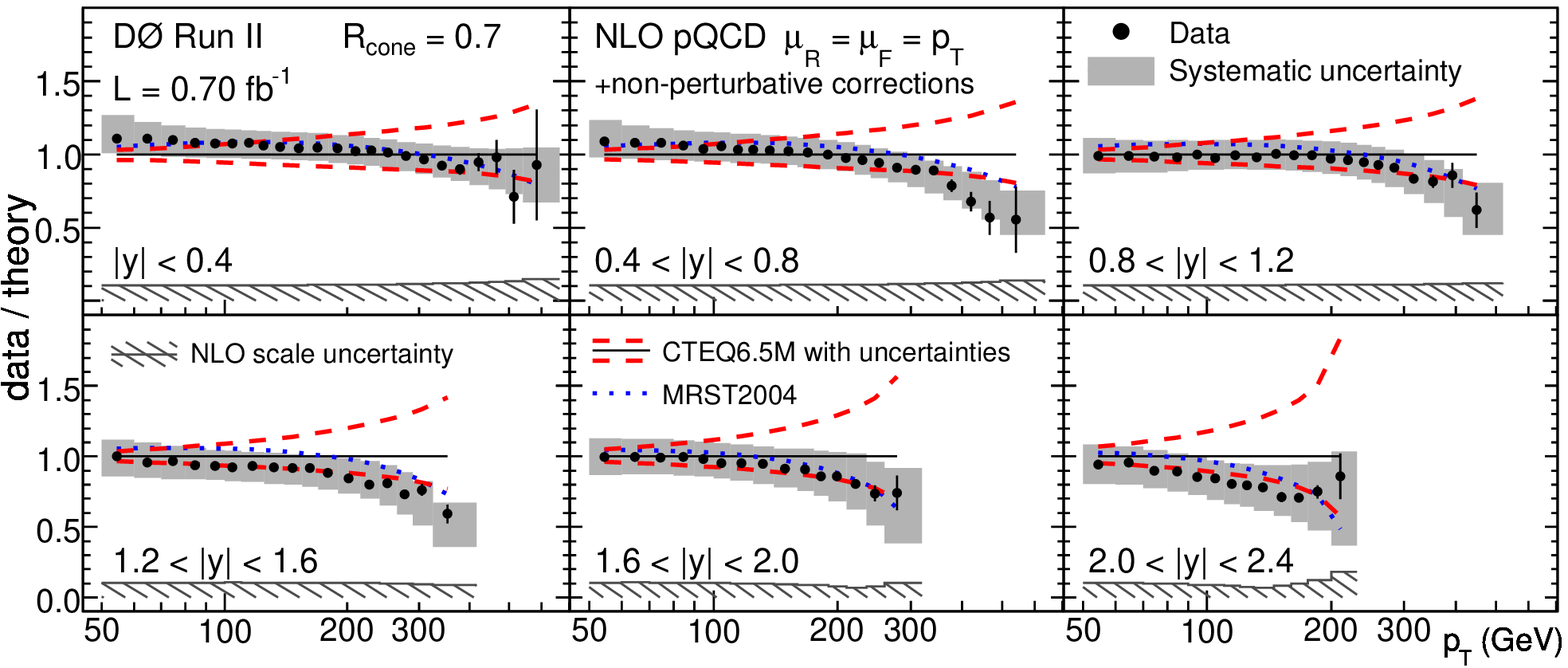}
\caption{{\em (top)} D{\O}'s observed inclusive jet differential cross sections
corrected to the particle level in six rapidity regions compared to
next to leading order (NLO) QCD predictions~\cite{D0-IncJets}.
The NLO QCD predictions are calculated 
with the CTEQ6.5M parton distribution functions;
{\em (bottom)} ratios of the measured cross sections
over the NLO QCD predictions. The data agrees with the theory quite well and 
has remarkably small systematic uncertainties.\label{Fig-D0-Incjets} }
\end{center}
\end{figure}
The \D0 collaboration analyzed 0.7 \fbinv  of data taken during $2004-2005$ to measure the inclusive jet cross section for
$\pT>50$ \GeVC in six rapidity bins, $|\Delta y| = 0.4$ wide, over the range $0 < |y| < 2.4$.
The data were collected by triggering on a  jet passing a \pT threshold. Six triggers with \pT thresholds 
of 15, 25, 45, 65, 95, and 125 \GeVC were used to collect data.  Due to high production rates, only predetermined fractions of the lower 
threshold triggers were recorded. The efficiency for triggering on jet events was measured using data collected with a muon trigger which
did not rely on calorimeter activity. These different jet triggers were combined to form the full \pT spectrum with
each trigger contributing to a unique \pT range. Only those data for which the trigger efficiency is $>98\%$ are used.
The events were required to have a reconstructed primary vertex and the position of the $p\bar p$ interaction be within 50
cm of the detector center along the beam direction.  This requirement ensured that the jets follow the projective geometry of 
the calorimeter and thus their energy was accurately measured. The consequence of this requirement was a reduction of only $7.0\pm0.5\%$ in the integrated luminosity.
The primary vertex was reconstructed using charged particle tracks measured using silicon micro-strip 
and scintillating fiber detectors located inside a solenoidal magnetic field of 2 T \cite{overview_d0}. 

The triggered data includes events containing
cosmic ray interactions, beam halo and detector noise. These contributions are mostly asymmetric and lead to
a large imbalance in the momentum in the plane transverse to beam direction, \Met. In contrast, for QCD jet production, 
\Met is ideally zero, apart from a small neutrino contribution. 
In QCD events, \Met arises mainly from fluctuations in calorimeter response and is much smaller than the total energy
observed in the detector. Most of these background events are removed by
requiring the ratio of \Met to the transverse momentum of the leading jet to be small. 
Remaining backgrounds are removed by requiring that the shape of energy deposition in 
the calorimeters be consistent with the expected shape from a hadronic jet.
The shape of energy deposition for a jet is very different from the energy deposited by a cosmic muon or a beam halo particle, as a jet
consists of many particles. These shape requirements also remove
photons and electrons.  These requirements are highly efficient for the signal and the remaining background is estimated to be
$<0.1\%$. 

The measured \pT of each jet is corrected for calorimeter non-linearity and energy lost in uninstrumented regions.
These average jet-by-jet energy corrections do not correct the smearing (bin-to-bin migration) of jets due to the finite energy resolution. 
This smearing is determined using an iterative procedure. It is assumed that the particle level physics (true) spectrum is described by
the function 
\begin{eqnarray}
F(\pT,y) = N_0 \left( \frac{\pT}{100 {\rm \, \GeVC}}\right)^\alpha \left( 1-\frac{2\pT \cosh(y_{\min})}{\sqrt{s}}\right)^\beta \exp(-\gamma\pT),
\end{eqnarray} 
where $y_{\min}$ is the rapidity lower bin edge. This functional form is a good representation of the NLO pQCD prediction and
fits the measured raw inclusive jet spectrum well.
This true spectrum is smeared using the jet energy resolution function, which is determined using $p\bar p$ collider dijet data and 
simulated dijet events.  
The resulting smeared spectrum is compared with data using a $\chi^2$ test. The process is iterated to determine the best 
parameters $(N_0, \alpha, \beta, \gamma)$ of the true function, $F(\pT,y)$.
This true spectrum is used to correct the migrations between bins in \pT in the observed data. 
In the central region, the migration correction is a multiplicative factor that is 
$0.8-0.9$ at low \pT and 0.7 at higher \pT, with a strong dependence on \y. The true spectrum $F(\pT,y)$ is measured separately for each rapidity bin.
The jet rapidity is measured very precisely and thus migration between rapidity bins is small. The $y$ migration corrections are less than
2\% in most bins and 10\% in the highest \pT bin where spectrum is the steepest. The rapidity unsmearing is applied after
the \pT unsmearing.
After the jet energy scale and resolution smearing corrections, the observed data distribution has been corrected to the particle-level jets and
is completely independent of the detector properties.  

The observed inclusive particle jet spectrum compared with the NLO pQCD predictions, corrected for underlying event and hadronization
effects, is shown in Figure~\ref{Fig-D0-Incjets}. The NLO pQCD predictions are calculated using NLOJet++ and FastNLO
\cite{NLOJet++,FastNLO} which use $\alpha_s^3$ matrix elements with $\muR=\muF=\pT^{\rm jet}.$
The parton distribution functions from CTEQ6.5M~\cite{CTEQ6.5M} are used, which include the Run I inclusive jet data.
The dashed curve shows the NLO pQCD prediction calculation using MRST 2004 divided by the same calculated using CTEQ6.5M.
The experimental uncertainty, dominated by the jet energy scale uncertainty, is  12\% at $\pT=50$ \GeVC and 17\% at $\pT=550$ \GeVC
in the $|y|<0.4$ bin. The uncertainty is higher in other \y bins. 
These results are the most precise to-date.
The main theoretical uncertainties arise from the
uncertainty on the PDFs and the missing higher order terms in the perturbation series. As is customary, the effect of
higher order terms is evaluated by varying the renormalization and factorization scale.
Fortunately, the change in cross section from varying these scales is almost independent of the \pT of the jets.
The predicted cross section changed by $\sim 10-15\%$ when the scale is changed to $\mu = 2\pT$ or $\mu = \pT/2$.
There is a good agreement between the data and the theoretical predictions over the
whole \pT range which spans 50 \GeVC to 550 \GeVC. Over this \pT range, the cross sections falls by 10 orders of magnitude. 

The data prefers the lower bound of the theoretical prediction, favoring a smaller gluon content of the proton at
high $x$. The theoretical
uncertainty arising from the uncertainties in the parton distribution functions is larger than the experimental uncertainties. 
These data, along with CDF inclusive jet data, have been used in the global fits to improve the precision of the gluon distribution function.  
The results of these fits are described  in Section~\ref{sec-pdfs}.

\subsubsection{CDF Collaboration\label{sec-CDF-Midpoint}}
The CDF collaboration measured the inclusive jet cross section for a cone size of $R=0.7$, using slightly more data, corresponding to 1.13\fbinv\cite{CDF-IncJet-Cone}.
The measurement spans five rapidity bins $|y|<0.1,\,0.1<|y|<0.7,0.7<|y|<1.1,\,1.1<|y|<1.6,\,1.6<|y|<2.1$. These bins
are matched to the CDF calorimeter structure~\cite{overview_cdf}
and thus are different than the binning used in \D0 analysis. The data was collected between $2002-2005$ using
jet triggers with four thresholds: 20, 50, 70 and 100 \GeVC. In order to not saturate the data acquisition system by jet triggers, 
only 1/808, 1/35, 1/8 of lower threshold triggers were recorded.
The transverse energy of each jet is corrected on average to form the jet \pT spectrum which was corrected for
bin-to-bin migration of jets due to finite jet energy resolution.
CDF used simulated events to evaluate the smearing corrections. The corrections depend on the shape of the true jet \pT spectrum 
and the jet energy resolution. A large sample of QCD jet events was generated using 
the \pythia event generator~\cite{pythia} and
passed through the CDF detector simulation. The detector simulation was tuned to describe the single particle response measured in 
$p\bar p$ collisions~\cite{CDF-JES}. These simulated data were analyzed using the same procedure as the one used for the real data
to obtained the smeared spectrum. The bin-to-bin migration effect was determined by taking the ratio of smeared jet spectrum and
the particle jet spectrum. For this procedure to be valid, the smeared \pT spectrum of the simulated events must match the
spectrum measured in data. The two spectra are very close but not exactly same.
The simulated particle jet smeared \pT spectrum was adjusted (re-weighted) to force it to agree with the measured spectrum.
Re-weighting changes the unsmearing corrections by only a few percent. The unsmearing correction is $<5\%$
for $\pT^{\rm jet}<300$ \GeVC and increases to as much as 20\% at $p_T=500$ \GeVC.

\begin{figure}[htbp]
\begin{center}
\includegraphics[width=0.45\hsize,clip]{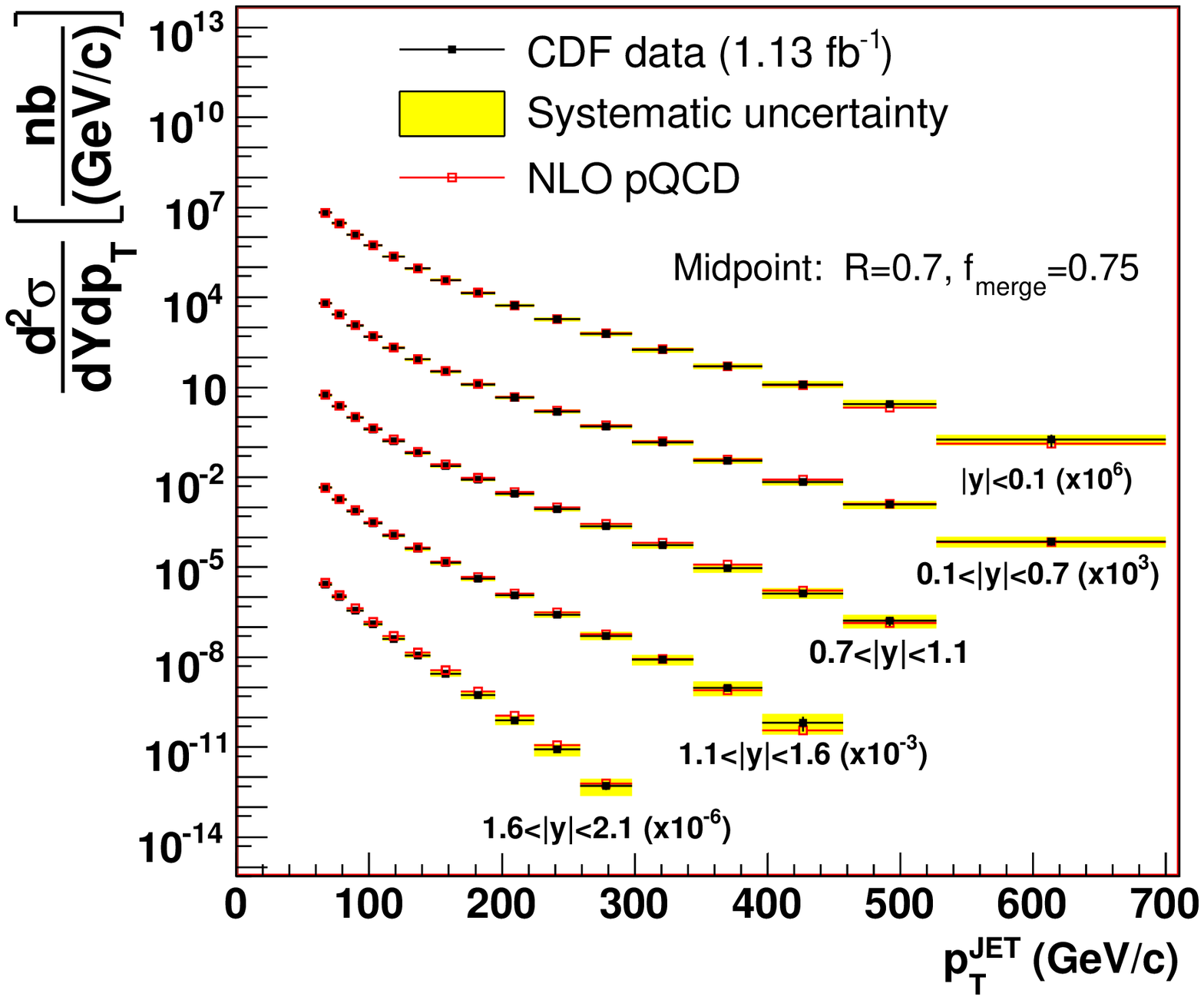}
\includegraphics[width=0.52\hsize,clip]{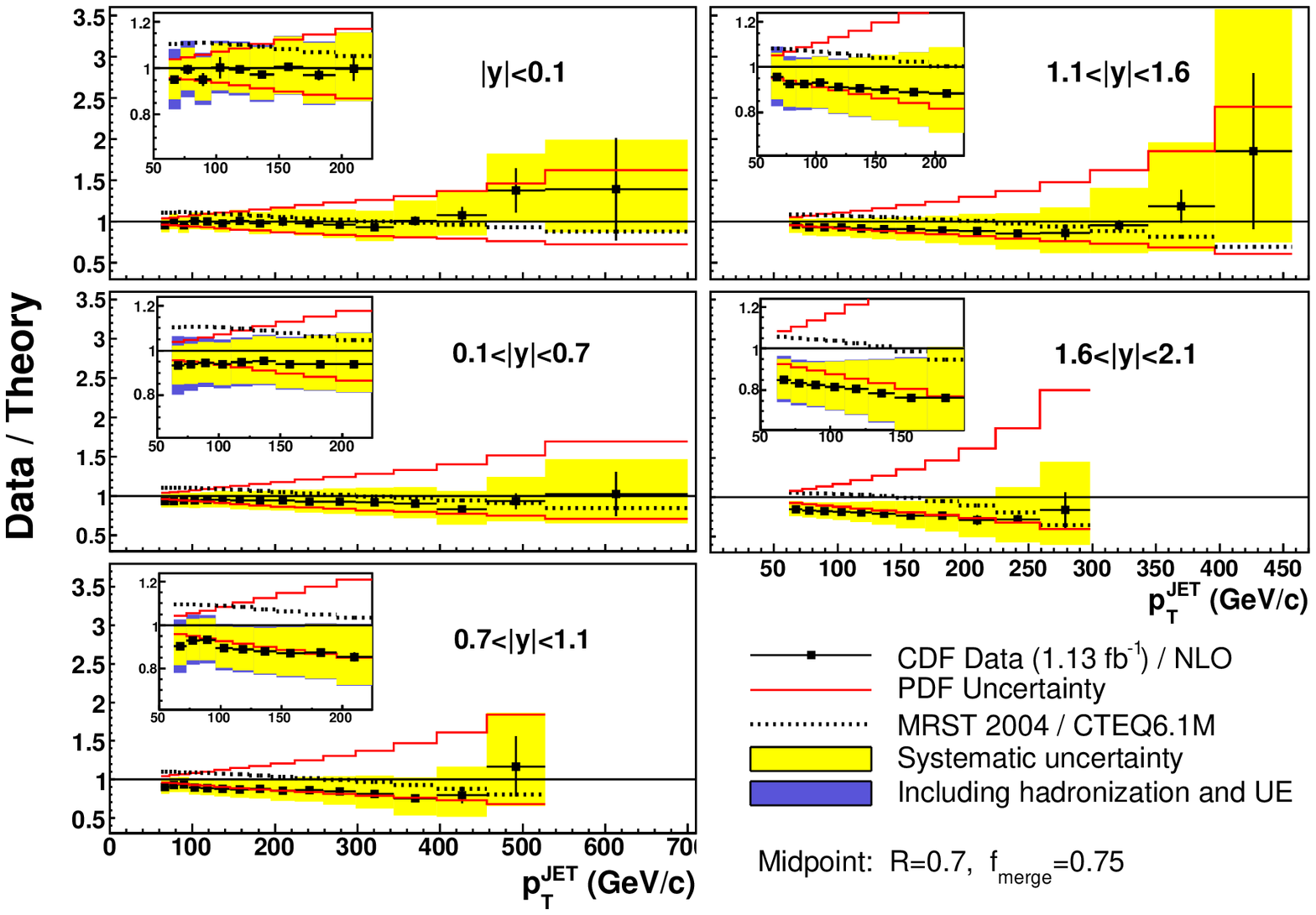}
\caption{{\em (left)} CDF's observed inclusive jet differential cross sections
corrected to the particle level in five rapidity regions compared to
next to leading order (NLO) QCD predictions~\cite{CDF-IncJet-Cone}.
The NLO QCD predictions are calculated 
with the CTEQ6.1M parton distribution functions.
{\em (right)} Ratios of the measured cross sections
over the NLO QCD predictions. The theory describes the data quite well.\label{Fig-CDFIncjetMidpoint}}
\end{center}
\end{figure}

The corrected jet \pT spectrum is compared to perturbative QCD predictions evaluated with the FastNLO~\cite{FastNLO}
program using the CTEQ6.1M parton distribution functions~\cite{CTEQ6.1M}. The renormalization and factorization scales ($\mu_R,\mu_F$) are chosen to be 
\pT/2, which are the same as used in the global QCD analysis to determine the PDFs~\cite{CTEQ6.1M}.
Using $\mu_R=\mu_F=\pT^{\rm jet}$ gives up to 10\% smaller predictions in the cross section. The perturbative QCD predictions are corrected for
underlying event and hadronization effects measured using the procedure described in Section~\ref{sec-pQCD}.
While clustering the partons produced by the FastNLO, CDF used an ad-hoc parameter $R_{sep}$ which was introduced to
mimic the split and merge procedure in iterative cone clustering algorithms~\cite{SteveEllisJetShapes}. At order $\alphas^3$, the
final state can have up to three partons. 
Depending on their relative \pT and their separation in $y-\phi$ space, these partons are clustered into two or three jets, 
Two partons are clustered into a single jet if they are within $R$ from the
jet centroid and within $R \times R_{sep}$ of each other. A value of $R_{sep}=1.3$ is used in this calculation.
An $R_{sep}=2.0$ (i.e. the midpoint algorithm without $R_{sep}$) yields less than a $5\%$ increase in cross section for NLO QCD predictions.
As shown in Figure~\ref{Fig-CDFIncjetMidpoint}, the data are in good agreement with the theoretical predictions. The
experimental uncertainty, dominated by the jet energy scale uncertainty, is comparable to the theoretical uncertainty which
is dominated by the PDF uncertainty.

\subsection{Measurement using the \kT  Clustering Algorithm\label{sec-kt-incjets}} 

The \kT clustering algorithm, which combines objects in pairs to reconstruct a jet, 
is infra-red and collinear safe at all orders in perturbation theory and is preferred over iterative cone algorithms.
The CDF collaboration has measured~\cite{CDF-IncJet-kt} the inclusive jet cross section using the \kT clustering algorithm 
with $D=0.4, 0.7,$ and \,1.0 using 1.0 \fbinv of data 
in the same rapidity region as used in the cone-based analysis~\cite{CDF-IncJet-Cone}. The analysis procedure is similar to the one described 
in Section~\ref{sec-CDF-Midpoint}, except the correction for multiple-interactions which are determined using a novel approach. 
The measured jet transverse momenta are corrected for this effect by removing a certain amount of transverse 
momentum, $\delta_{\pT}^{mi} \times(N_V-1)$
where $N_V$ denotes the number of primary vertices in the event.
The value of \pileup was determined by requiring the shape of the \pT spectrum at high instantaneous luminosity (\instlum) to be the 
same as the one at the low \instlum. After making the shape of two spectra the same, the data at low \instlum and high \instlum are combined.
The study was carried out
independently for each rapidity region and the results were consistent with a common value $\pileup= 1.86 \pm 0.23$ \GeVC.
The corresponding correction for the cone jets is $0.97\pm 0.29$ \GeVC, which is measured by summing the \pT in a cone of $R=0.7$
in minimum bias events. 

\begin{figure}[htbp]
\begin{center}
\includegraphics[width=0.450\hsize]{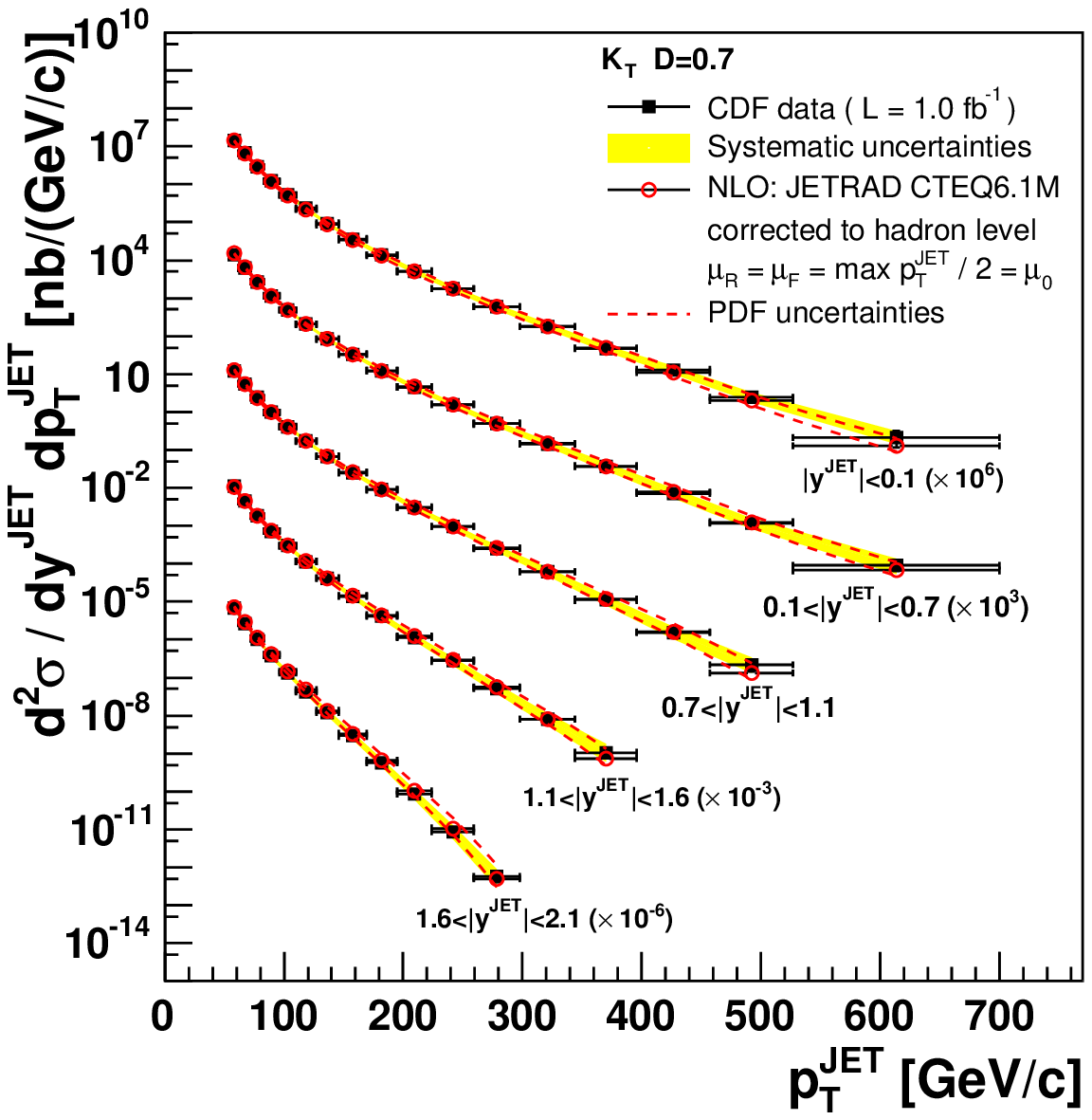} \\
\includegraphics[width=0.450\hsize]{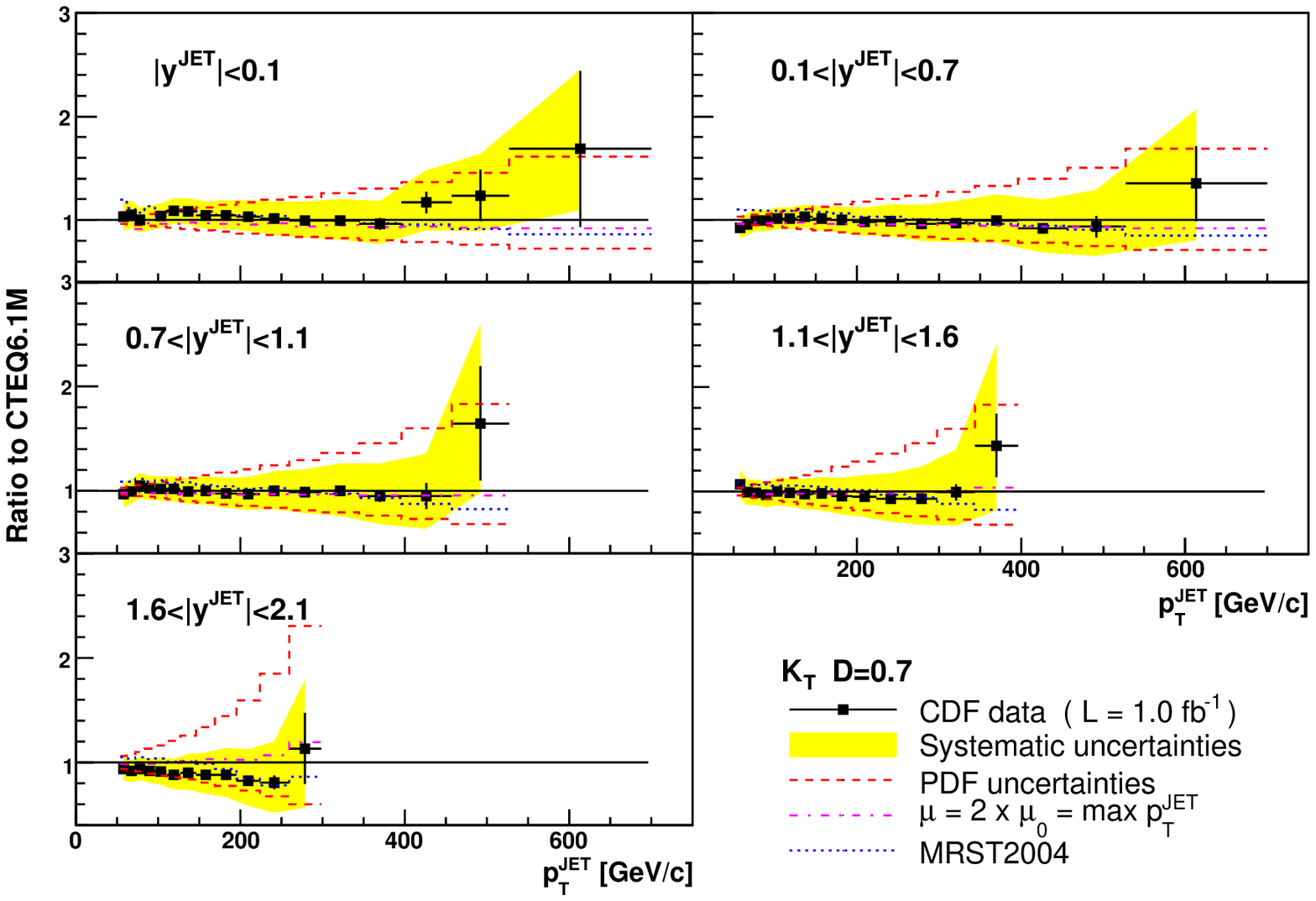}
\caption{{\em (top)} CDF's inclusive jet differential cross sections measured using the \kT clustering 
algorithm in five rapidity regions.
The data has been corrected to the particle jets and is compared to
next to leading order (NLO) QCD predictions~\cite{CDF-IncJet-kt}.
The NLO QCD predictions are calculated 
with the CTEQ6.1M parton distribution functions.
{\em (bottom)} Ratios of the measured cross sections
over the NLO QCD predictions. The perturbative NLO QCD calculations are in good agreement with the data.\label{Fig-CDF-IncjetKt}}
\end{center}
\end{figure}

The \pT spectra are compared in Figure~\ref{Fig-CDF-IncjetKt} with NLO QCD predictions
using the CTEQ6.1M PDFs~\cite{CTEQ6.1M}  with $\mu = 0.5 \times \pT^{\rm max jet}$ for $D=0.7$. The theoretical predictions are calculated
using the JETRAD~\cite{JETRAD} program. 
The data are in very good agreement with QCD predictions except in the highest rapidity bin $(1.6<|\y|<2.1)$, where the data are
lower than the prediction, but well within experimental systematic uncertainties. The theoretical uncertainties are dominated by
the PDF uncertainties and are comparable or larger than the experimental uncertainties. The theoretical predictions using the MRST2004 PDFs is 
very close to those based on the CTEQ6.1M PDFs, except in the $1.6<|y|<2.1$ bin, where the MRST2004 cross section is smaller, 
but within the PDF uncertainty on the CTEQ6.1M prediction.  The results for jet size $D=0.4$ and $D=1.0$ show similar behavior.

The jet \pT spectra measured using two different clustering algorithms are expected to be different and can be compared 
only via theoretical predictions. The ratios of data/theory from two analyses were compared and the 
two ratios were in very good agreement with each other 
except in the $0.7<|y|<1.1$ region where the \kT cross section is $\sim 5\%$ higher.
In this \y region, the CDF calorimeter coverage is not uniform which leads to a large variation in calorimeter response and poor jet energy resolution. 
The two CDF analyses have similar experimental uncertainties. Thus one concludes that both the \kT and the cone clustering algorithms can be successfully used
at the hadron colliders.

\subsection{Determination of Gluon Distribution Function\label{sec-pdfs}}
The parton distribution function (PDF) $f_i(x,\mu)$, which is the probability to find a parton with a type $i=g,q,\bar{q}$ with 
momentum fraction $x$ and mass scale $\mu$, must be experimentally determined. The PDFs for gluons and light quarks 
and anti-quarks ($u,d,s,\bar{u},\bar{d},\bar{s}$)  are normally determined from experimental data. For heavier quarks, i.e. $c$ and $b$, 
they are normally dynamically generated through gluon splitting. Data from $e^{\pm}p$ collisions at the ZEUS and H1 experiments~\cite{hera_qcd}, $\nu p$, $\bar{\nu} p$, $\nu n$,
$\bar{\nu} n$ collisions at CCFR/NuTeV~\cite{nutev}, and 
Drell-Yan (lepton pairs and $W/Z$ bosons) production in $pp$~\cite{drell-yan} and $p\bar{p}$ collisions~\cite{tevatron_drell_yan},
jet data from Tevatron and data from many other experiments, especially low energy, are used to extract the PDFs using a global fit.
These experiments are sensitive to different $f_i$. For example, $e^{\pm}p$ experiments are sensitive to the sum of the $q$ and $\bar{q}$ distributions
weighted by $e^2_{q(\bar q)}$ and can not distinguish between quark and anti-quark distributions. 
Neutrino and anti-neutrino data are used to differentiate between $q$ and $\bar{q}$.
The Tevatron jet data play a significant role in constraining the gluon distribution at large $x$. 
The gluon distribution at low $x$ are mainly determined from the scaling violations in lepton-nucleon scattering data.
Normally, the results of these global analysis are fit at some initial scale $\mu_0$ using tens of parameters.
The PDF can be evolved to any arbitrary scale $\mu$ using QCD evolution equations which are 
available at NNLO approximation in perturbation theory. 

\begin{figure}[htbp]
\begin{center}

\includegraphics[width=0.450\hsize,clip]{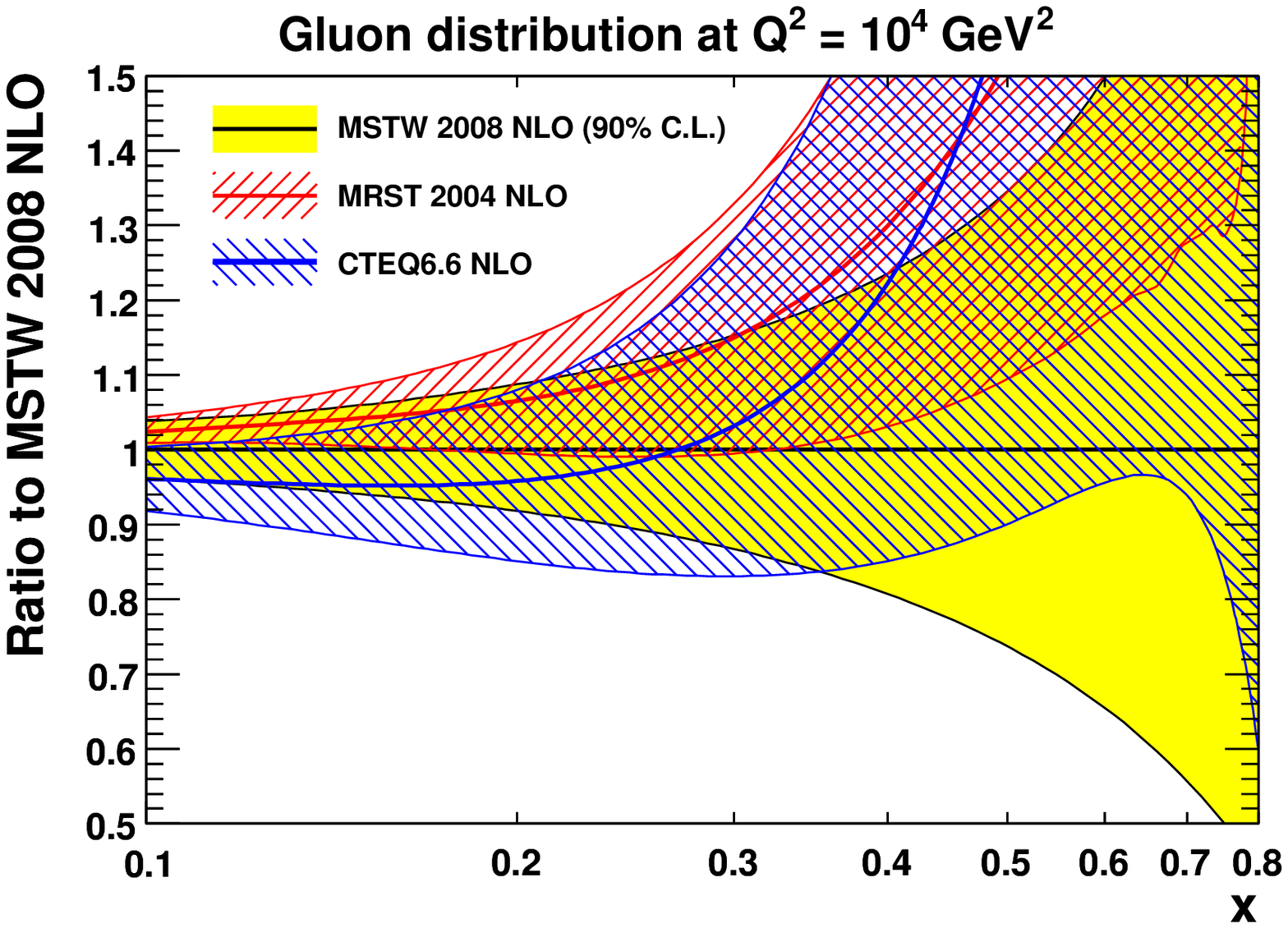}
\includegraphics[width=0.450\hsize,clip]{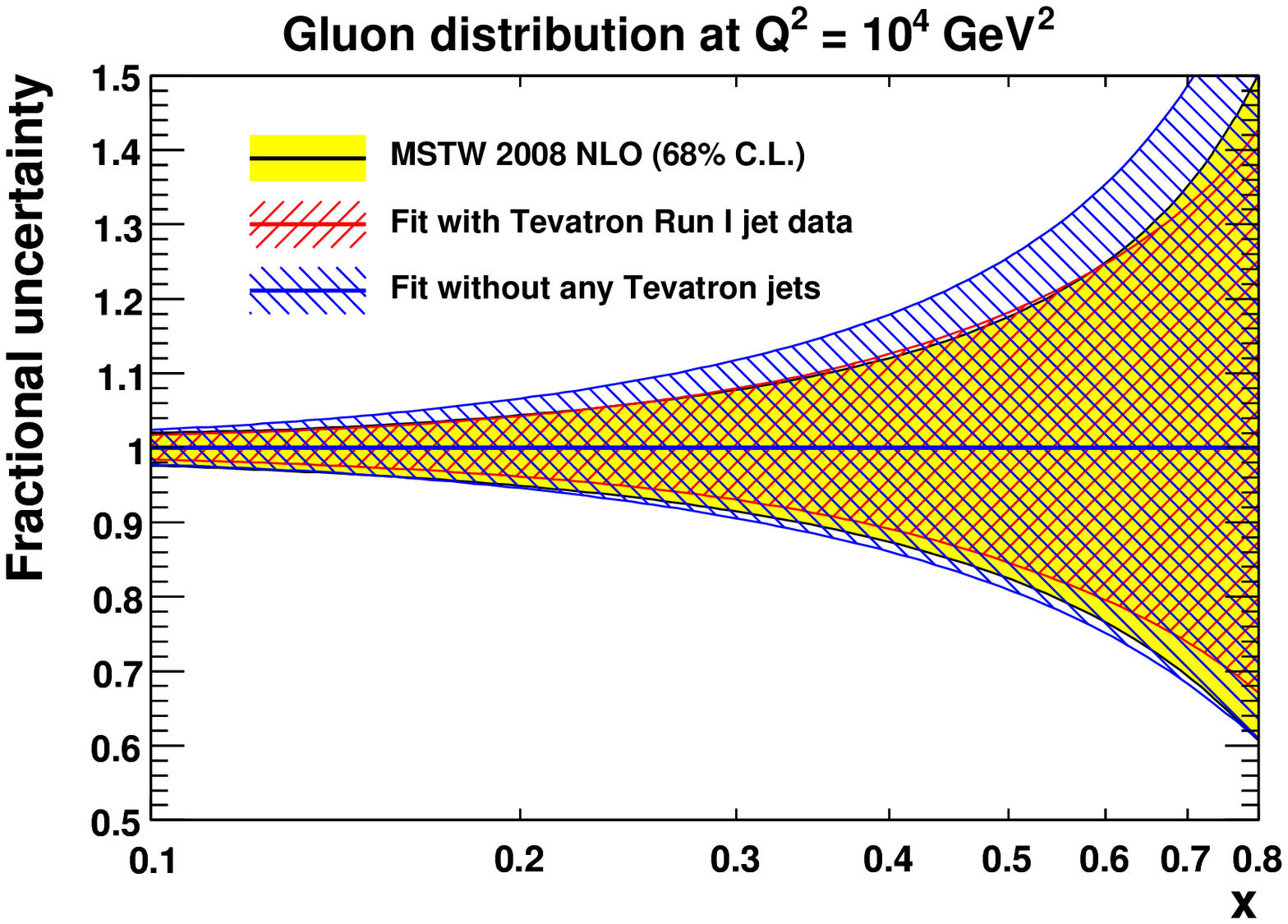}

\caption{
{\em (left)} Ratios of various gluon PDFs to the MSTW2008 PDF.
{\em (right)} The MSTW2008 gluon PDF uncertainties for variations on the inclusion or exclusion of Tevatron data~\cite{MSTW}.
\label{Fig-MSTW-errors}}
\end{center}
\end{figure}

Both the MSTW~\cite{MSTW} and  CTEQ~\cite{CTEQ-IncJet-2009} collaborations have included Run II jet data in their global analysis
to update the PDFs. The three Run II inclusive jet
measurements are more accurate than Run I measurements, span a larger \pT range and are consistent with each other~\cite{MSTW}.
As two CDF measurements used the same data, the MSTW collaboration decided to use the \kT jet spectrum whereas the CTEQ collaboration is using
the cone-based measurements. The Run I jet measurements do not play a significant role in the fit and thus the MSTW collaboration
has dropped those data from the new fits. Comparisons of the gluon distribution $g(x)$ determined in the new MSTW fit with the gluon distributions 
from MRST2004~\cite{MRST2004} and CTEQ6.6~\cite{CTEQ6.6} fits along with the 90\% uncertainty band are shown in 
Figure~\ref{Fig-MSTW-errors} (left) for $\mu^2=Q^2=10^4$ ${\rm GeV}^2$. The new $g(x)$ is 
lower than previous fits for $x\ge 0.3$ but within the still large
systematic uncertainties. As $\alphas$ and $g(x)$ always appear as a product, the values of \alphas and $g(x)$ are strongly correlated.
The value of \alphas in the three sets of PDFs is different and thus $g(x)$ is also expected to be slightly different. 
The fractional uncertainty on the gluon distribution is shown in Figure~\ref{Fig-MSTW-errors} (right).
At $x=0.4$ and $\mu^2 = 10^4$ GeV$^2$, the uncertainty reduces from 18\% when the jet data are excluded from the fit 
to 12\% when jet data are included in the fit. This modest extra constraint  will make the predictions more precise at the LHC
in processes where gluon-quark scattering dominates.  

\subsection{Determination of the strong coupling constant}
The hard cross section for jet production ($\widehat{\sigma}_{ij}$) in Equation~(\ref{eqn-hard_scatter}) at large \pT can 
be expanded in powers of the strong coupling constant to the $n^{\rm th}$ order in perturbation theory 
\begin{eqnarray}
\widehat{\sigma}_{ij} = \alphas^2 \sum_{m=0}^{n} c_{ij}^{(m)}\alphas^m.
\label{eqn-alpha_s}
\end{eqnarray}
where the perturbative coefficients $c_{ij}^{(m)}$ are functions of the kinematic variables and the factorization and renormalization
scales only. The coefficients $c_{ij}^{(m)}$ are 
available for  $m = 0$ and $m = 1$. Using the jet cross section measurement, the 
strong coupling constant \alphas can be determined using Equation~\ref{eqn-hard_scatter} provided the parton distributions functions are known.
This technique was used by the CDF collaboration to measure \alphas at different \pT values and 
show its running with the hard scattering scale in Run I~\cite{CDF-alphas}. In Run II, the \D0 collaboration has used the same principle but an improved
technique to measure \alphas~\cite{D0-alphas} from the data used to measure the inclusive jet cross section. 
The hadronization and underlying event corrections were determined  using \pythia and applied to theory predictions.
The hadronization (underlying event) corrections
vary between -15\% (+30\%) to -3\% (+6\%), for \pT = 50 \GeVC to 600 \GeVC~\cite{D0-IncJets}.

The perturbative results are the sum of ${\cal O}(\alphas^3)$ pQCD calculation~\cite{NLOJet++,FastNLO}, 
supplemented with ${\cal O}(\alphas^4)$(2-loop) corrections for the threshold effects~\cite{QCD-2-loop}. The PDFs are taken from the MSTW2008
next-to-next-to-leading order (NNLO) parametrization and the renormalization and factorization \muR,\,\muF scales are set equal to the \pT of the jet.
The NNLO (NLO) MSTW2008 PDFs are available for 21 different values of $\alpha_s(M_Z)$ ranging from 0.107 - 0.117 (0.110 - 0.130) in steps of 0.001. This \alphas is used
to evolve PDFs.

Commonly available parton distributions from the MSTW and CTEQ groups include Tevatron jet data in the global fit. 
To avoid any correlation between the input PDFs and
the extracted \alphas, only 22 of 110 available jet data points are used. These selected points contribute to the $x$ region ($x\le 0.25$) where PDFs in global fits
are mainly determined by other experimental data and are not strongly influenced by Tevatron jet data. 
The jet data starts to affect $g(x)$ at $x \sim 0.2$. The change in $g(x)$ due to inclusion of jet data is less than 5\% for $x\lesssim 0.25$~\cite{MSTW}. 

\begin{figure}[htbp]
\begin{center}
\includegraphics[width=0.450\hsize]{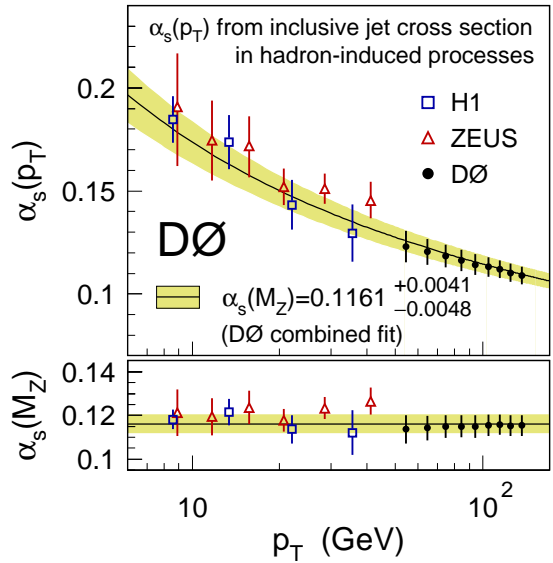}
\caption{$\alpha_s$ measurement as a function of the transverse momentum of the jet from several experiments, as compared to the expected variation of $\alpha_s$, setting $\alpha_s(M_Z) = 0.1184 \pm 0.0007$~\cite{D0-alphas}.
\label{Fig-d0-alphas}}
\end{center}
\end{figure}

The central $\alphas(M_Z)$ result is obtained by minimizing $\chi^2$ with respect to $\alphas(M_Z)$ and integrating over the nuisance parameters for the 
correlated uncertainties.  The variation of $\alphas(\pT)$ vs \pT is shown in Figure~\ref{Fig-d0-alphas} (top). 
The running of \alphas as a function of jet \pT follows the QCD evolution equation.
The data points from the H1 and ZEUS experiments follow the same curve but have large uncertainties.
The \alphas, evolved to $\mu=M_Z$, is shown in Figure~\ref{Fig-d0-alphas} (bottom). The combined result of 22 selected points 
$\alphas(M_Z)= 0.1161^{+0.0041}_{-0.0048}$ is 
consistent with the world average $0.1184\pm 0.0007$~\cite{world-alphas} although the
uncertainties, mostly theoretical, are large.

\section{Search for Physics Beyond the Standard Model using Jets}
Due to large theoretical and experimental uncertainties in the jet production rate, the high \pT tail cannot be used 
to improve upon the current limits on new interactions \cite{compositeness}.  However, there are 
kinematic distributions from jet events that can be used to identify physics beyond the Standard Model.
Below we describe these searches using the dijet mass spectrum and the dijet angular distributions. These searches are
not very sensitive to the jet energy scale, parton distribution functions or renormalization or factorization scales.

\subsection{Dijet Mass Spectrum} 
Many new physics models predict particles which decay into two high \pT jets.  These particles
can be identified by the reconstructed mass of the dijet system provided their intrinsic mass width
is narrow. Such models include excited quarks~\cite{excitedquarks},
axigluons~\cite{axigluons}, flavor-universal colorons~\cite{coloron}, color-octet techni-$\rho$~\cite{techni-rho},
Randall Sundrum (RS) gravitons~\cite{RS-graviton}, heavy vector bosons~\cite{WZprime} 
and diquarks in the string-inspired $E_6$ model~\cite{diquark}. 
The excited quarks \qstar decay into $qg$. Heavy vector bosons $W^\prime$, $Z^\prime$ decay into $q\bar{q}$ or  $q\bar{q}^\prime$.
The axigluon $A$ decays into $q\bar q$, and $E_6$ diquarks $D\,(D^c)$ decay into $\bar{q}\bar{q}(qq)$.
The RS Graviton \RSGraviton and color-octet techni-$\rho$ \rhoT both decay into either a $qq$ or $gg$ pair but their
branching ratios are different.

\begin{figure}[h]
\begin{center}
\includegraphics[width=0.450\hsize]{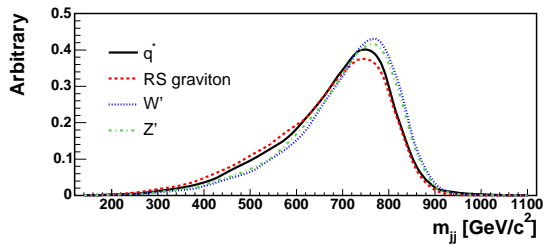}
\caption{Expected dijet mass distributions for simulated signals for the following new physical models: $q^*\rightarrow qg$, RS graviton ($\rightarrow gg,\,q\bar q$)
and $W^\prime\rightarrow q\bar{q}$ and $Z^\prime\rightarrow q\bar{q}$ with a mass of 800 \GeVCsqr.}
\label{Fig-MassTemplates}
\end{center}
\end{figure}

All these models predict an intrinsic mass width which is much smaller than both the detector resolution and mass broadening effects due 
to QCD radiation. 
These models can be divided into three categories depending on the decay channel, i.e. $gg$, $gq$ and $qq$.
The expected mass shapes for \qstar, \RSGraviton, \Wprime, and \Zprime particles with a mass of 800 \GeVCsqr are shown in 
Figure~\ref{Fig-MassTemplates}. Because \qstar and \RSGraviton decay into gluons, their widths are broader than the widths for \Wprime and \Zprime. 
Gluons radiate more than quarks, resulting in a broader dijet mass distribution. These distributions are close and change the final limits 
by only 10-20\%. These shapes can be used to search for resonance structure, independent of the model details.

\begin{figure}[htbp]
\begin{center}
\includegraphics[width=0.450\hsize]{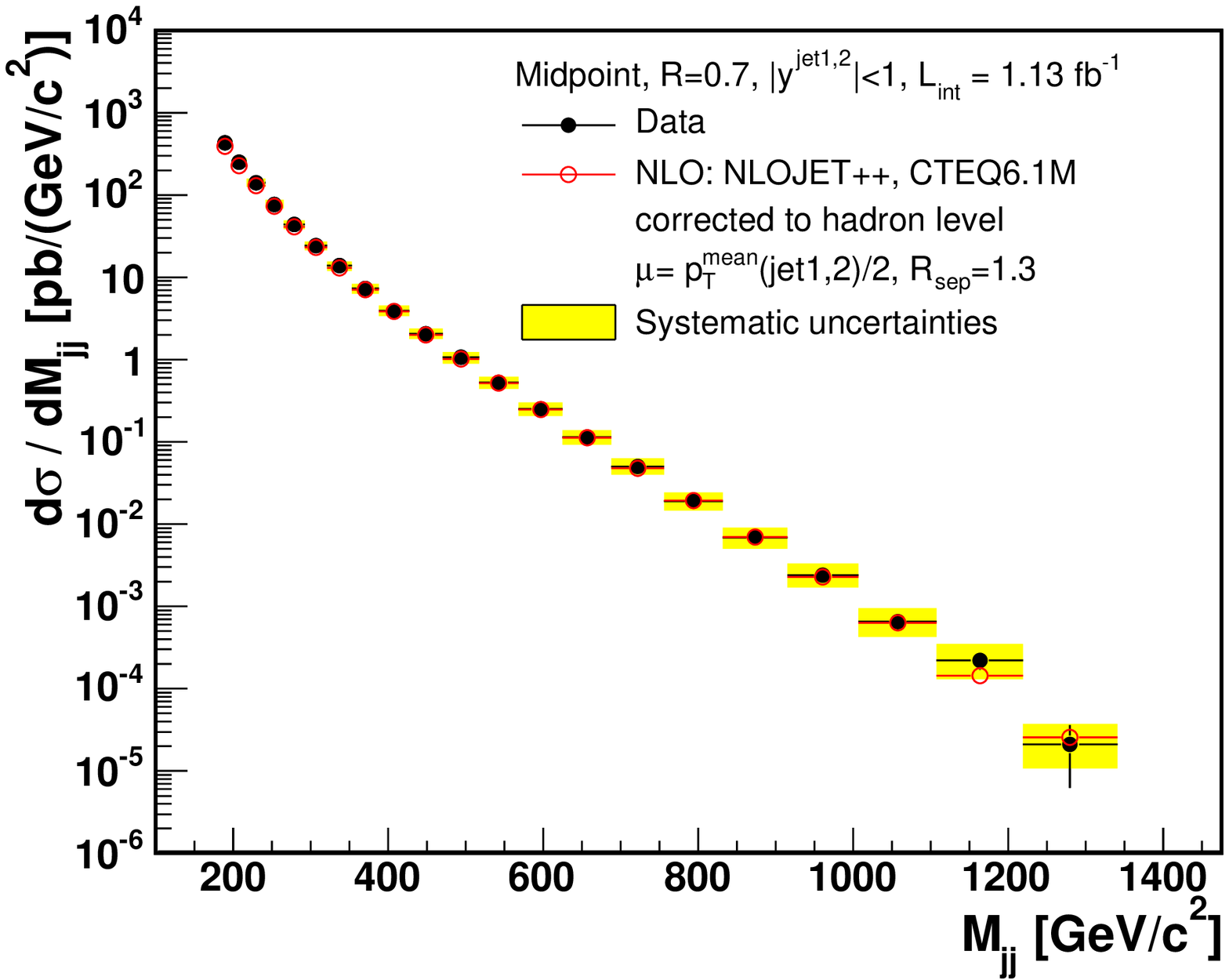}
\includegraphics[width=0.450\hsize]{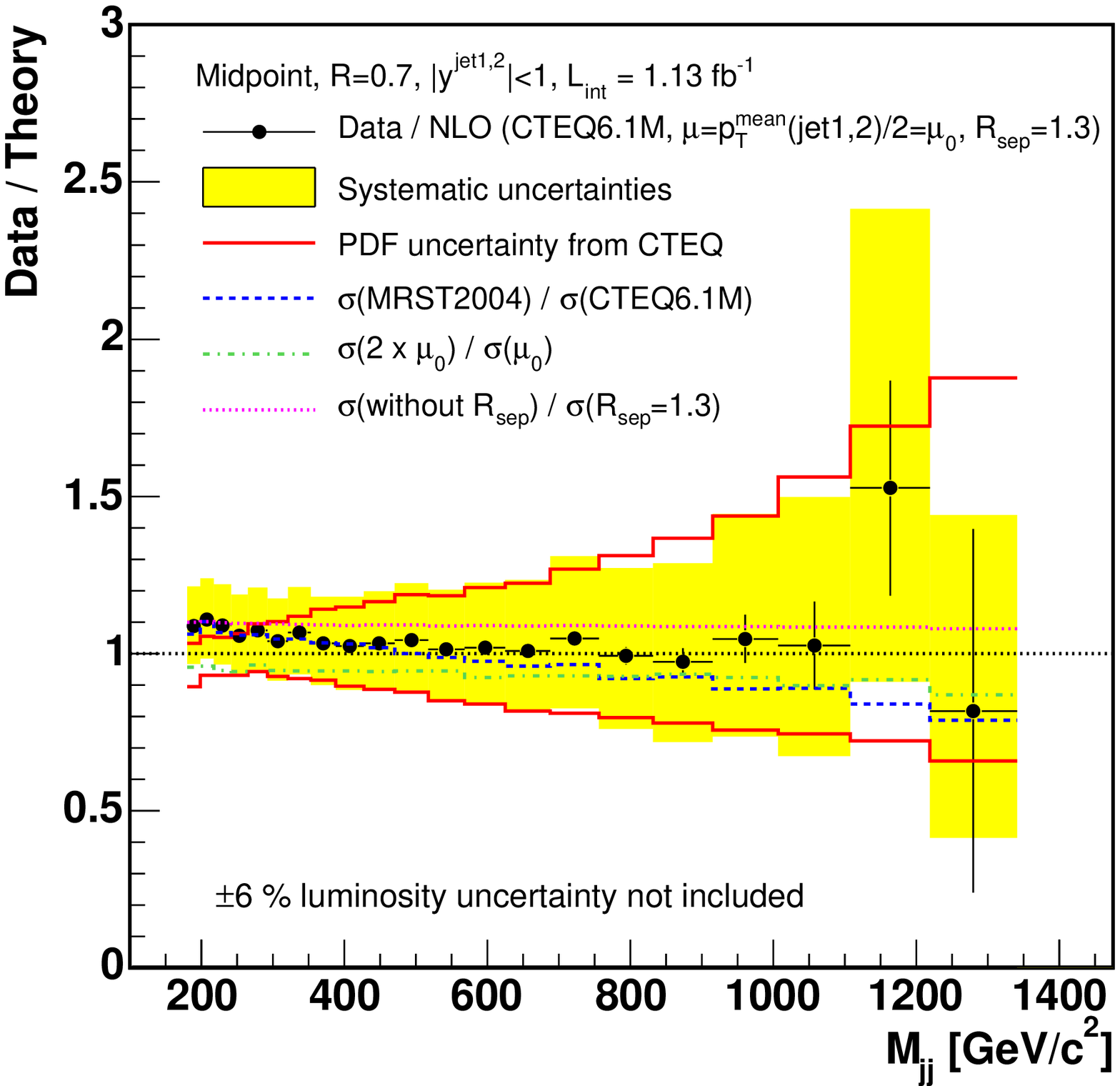}
\caption{{\em (left)} CDF's measurement of the dijet mass cross section for events in which the two highest \pT have $|y|< 1.0$~\cite{cdf_dijet_mass} compared to NLO calculations using CTEQ6.1M.
{\em (right)} Ratio of data to NLO theory.  The experimental uncertainties are dominated by jet energy scale uncertainty and are comparable to the theoretical PDF uncertainty. \label{Fig-CDF-dijets}}
\end{center}
\end{figure}

\begin{figure}[h]
\begin{center}
\includegraphics[width=0.450\hsize]{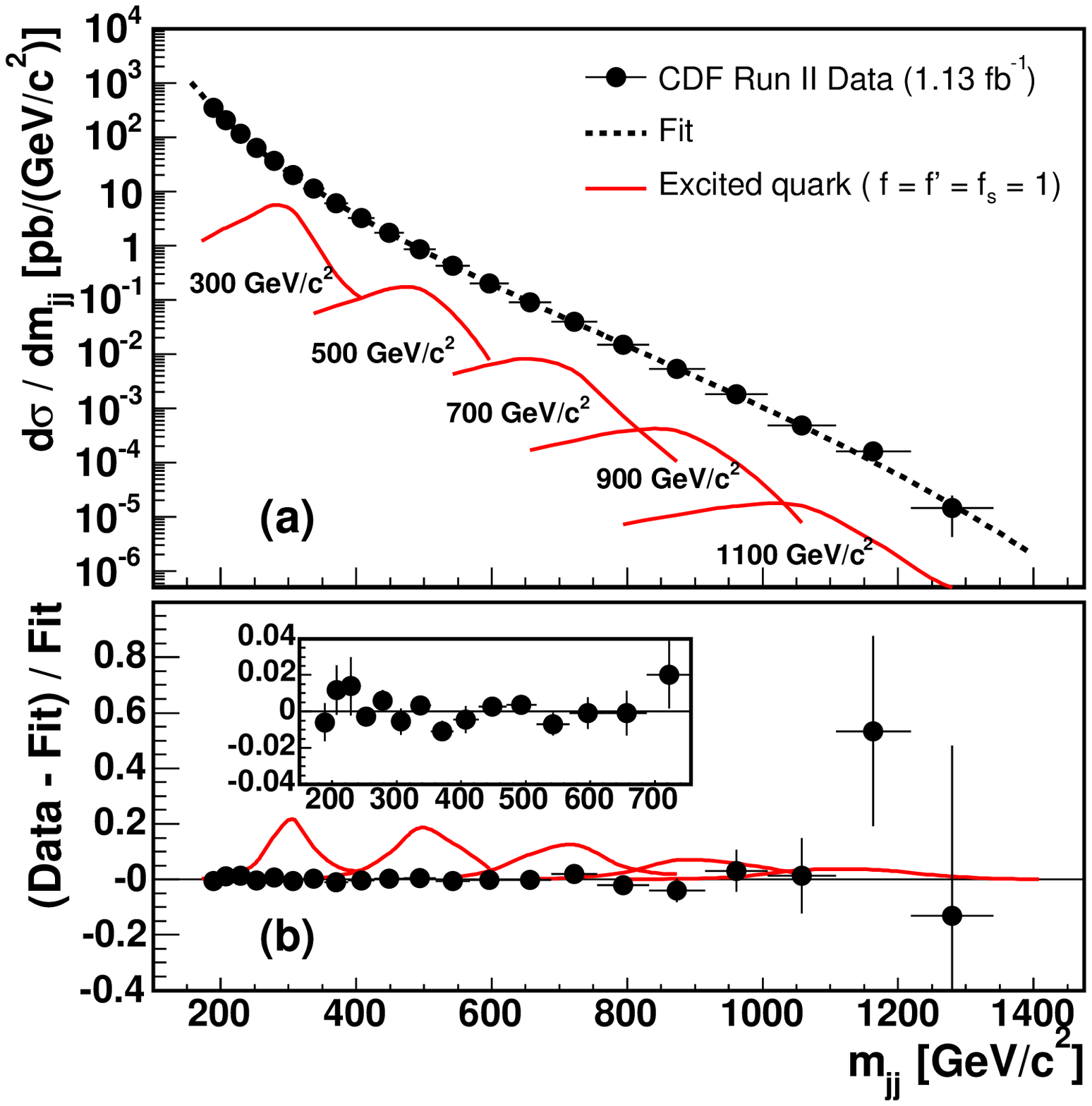}
\caption{(a) CDF's  measured dijet mass spectrum~\cite{cdf_dijet_mass}. The dashed curve shows 
the fit to Equation~\ref{eqn-CDF-dijets}. Also shown are the predicted dijet mass distributions
of the excited quark $q^*$ for the masses of 300, 500, 700, 900 and 1100 \GeVCsqr
respectively. (b) The fractional difference between the measured dijet mass difference
and the fit (points) compared to $q^*$ signals divided by the fit to the measured dijet
mass spectrum (curves).
\label{Fig-CDF-dijets-resonance}}
\end{center}
\end{figure}

To measure the dijet mass spectrum~\cite{cdf_dijet_mass},
the CDF collaboration used the same data set as used in the inclusive jet cross section measurement described 
in Section~\ref{sec-CDF-Midpoint}~\cite{CDF-IncJet-Cone}.
The dijet mass is reconstructed from the two highest \pT jets using
\begin{eqnarray}
m_{jj} = \sqrt{(E_1+E_2)^2- (\vec{p}_1+\vec{p}_2)^2}.
\end{eqnarray}
Jets produced by new physics 
are expected to be produced more centrally
than by Standard Model processes and we expect a better signal to background ratio in 
the central region.  Thus only those events in which two leading jets have $|y|\le 1.0$ are used.
Moreover, the CDF calorimeter is best understood in this region.
As shown in Figure~\ref{Fig-CDF-dijets}, these data, after all corrections, are in good agreement with the NLO QCD calculations.
The dijet mass spectrum before smearing corrections is shown in Figure~\ref{Fig-CDF-dijets-resonance}.
Smearing corrections are not used to avoid any degradation of the resonant structure, if any, in the data.
The measured mass spectrum is fit to a smooth background given by 
\begin{eqnarray}
\frac{d\sigma}{dm_{jj}} = {p_0(1-x)^{p_1}}/{x^{p_2+p_3\ln(x)}},   \hspace{5ex} x={m_{jj}}/{\sqrt{s}},
\label{eqn-CDF-dijets}
\end{eqnarray}
where $p_0,p_1,p_2,$ and $p_3$ are free parameters.
The dijet mass spectra predicted by \pythia, \herwig, and NLO pQCD
can be described well by this functional form. 
The fit to the measured  dijet mass spectrum is shown in 
Figure~\ref{Fig-CDF-dijets-resonance}(a). The data are well described by this smooth function with a $\chi^2$ of 16 for 17 degrees of freedom.
The deviation from the smooth curve is shown in Figure~\ref{Fig-CDF-dijets-resonance}(b).
These data are used to determine the exclusion limits
on the existence of new particles decaying into jets,
as there is no evidence for the existence of any resonant structure. 
The experimental limits are determined for the $\sigma^{sig} \equiv \sigma\cdot B\cdot A$, where
$\sigma$ is the theoretical new particle production rate, $B$ is probability of its decaying into
two jets and $A$ is the kinematic acceptance of the resulting particle jets to have $|y|<1.0$. 

The upper limits on $\sigma^{sig}$ are evaluating using a likelihood function 
\begin{eqnarray}
L=\prod_i \mu_i^{n_i} \exp(-\mu_i)/n_i!
\end{eqnarray}
where $\mu_i = n_i^{sig} + n_i^{QCD}$ is the predicted number of events in bin $i$. The QCD dijet background $n_i^{QCD}$ is
determined using Equation~(\ref{eqn-CDF-dijets}) by evaluating  ${\cal L}_i\cdot \epsilon_{i} \cdot \Delta m_{jj}\cdot d\sigma/dm_{jj}|_i$
where $\Delta m_{jj}$, $\epsilon_{i}$, and ${\cal L}_i$, are the bin width, the trigger efficiency and the integrated luminosity for
bin $i$ respectively.
The expected signal events $n_i^{sig}$ is given by $\sigma^{sig}\cdot {\cal L}_i\cdot \epsilon_i\cdot (n_i/n_{tot})$ 
where $\epsilon_i$ is the signal event selection efficiency in the $i^{\rm th}$ dijet mass bin and  $n_i/n_{tot}$ is the predicted
signal fraction in bin $i$. For each value of $\sigma^{sig}$, the likelihood is maximized with respect to
the four parameters in Equation~(\ref{eqn-CDF-dijets}).  This profiled likelihood is integrated over  
Bayesian priors for the parameters describing the systematic uncertainties~\cite{Luc-Bayesian}.
A  flat prior on $\sigma^{sig}$ is used to extract a Bayesian upper limit on that parameter.

The mass exclusion limits for \Wprime, \Zprime, \qstar and \RSGraviton are determined by comparing
the limits obtained using their respective signal shapes and the predicted theoretical cross section.
For other models, the limits obtained for the above four signal shapes are used. The \qstar signal shape ($qg$) is 
used for axigluons, the flavor-universal coloron and the $E_6$ diquark, as these particles do not decay into modes
which include gluons and thus their signal shapes are expected to be narrower than the \qstar signal shape. For
\rhoT, the limits obtained for the \RSGraviton shape are used. Both the \rhoT and the \RSGraviton decay into $q\bar{q}$ or $gg$, 
but the branching fraction of \RSGraviton into $gg$ is higher. Thus in all the above cases, the obtained exclusion limits
are conservative. The limits for these models are given in Table~\ref{Table-CDF-dijet-limits}.

\begin{table}[h]
\begin{center}
\caption{\label{Table-CDF-dijet-limits}The limits on the masses of particles decaying into dijet in various models.  
Limits are in units of \GeVCsqr~\cite{cdf_dijet_mass}.}
\begin{tabular}{lcc}
\hline
\hline
Model        & Parameters        &  Excluded Region \\
\hline
\qstar       & $f=f^\prime=f_s$  & 260 - 870 \\
axigluons    &                   & 260 - 1250 \\
coloron      &                   & 290 - 630 \\
$E_6$ diquark&                   & 260 - 1100 \\
\rhoT        &                   & 280 - 840  \\
\Wprime      & SM                & 280 - 840 \\
\Zprime      & SM                & 320 - 740 \\
\RSGraviton  & $k/M_{Pl}=0.1$    &     -   \\
\hline
\hline
\end{tabular}
\end{center}
\end{table}

\begin{figure}[htbp]
\begin{center}
\includegraphics[width=0.450\hsize]{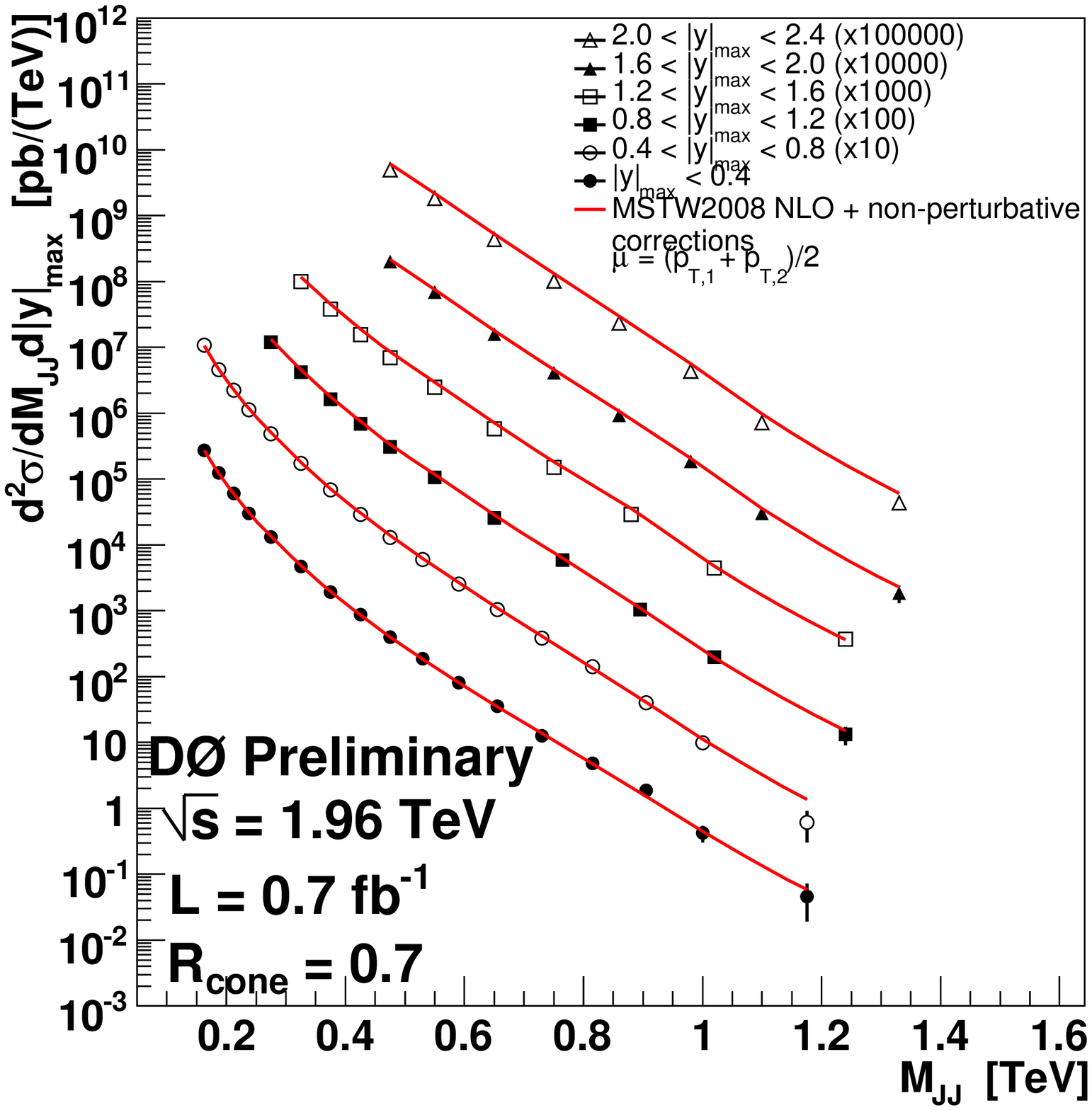}
\includegraphics[width=0.450\hsize]{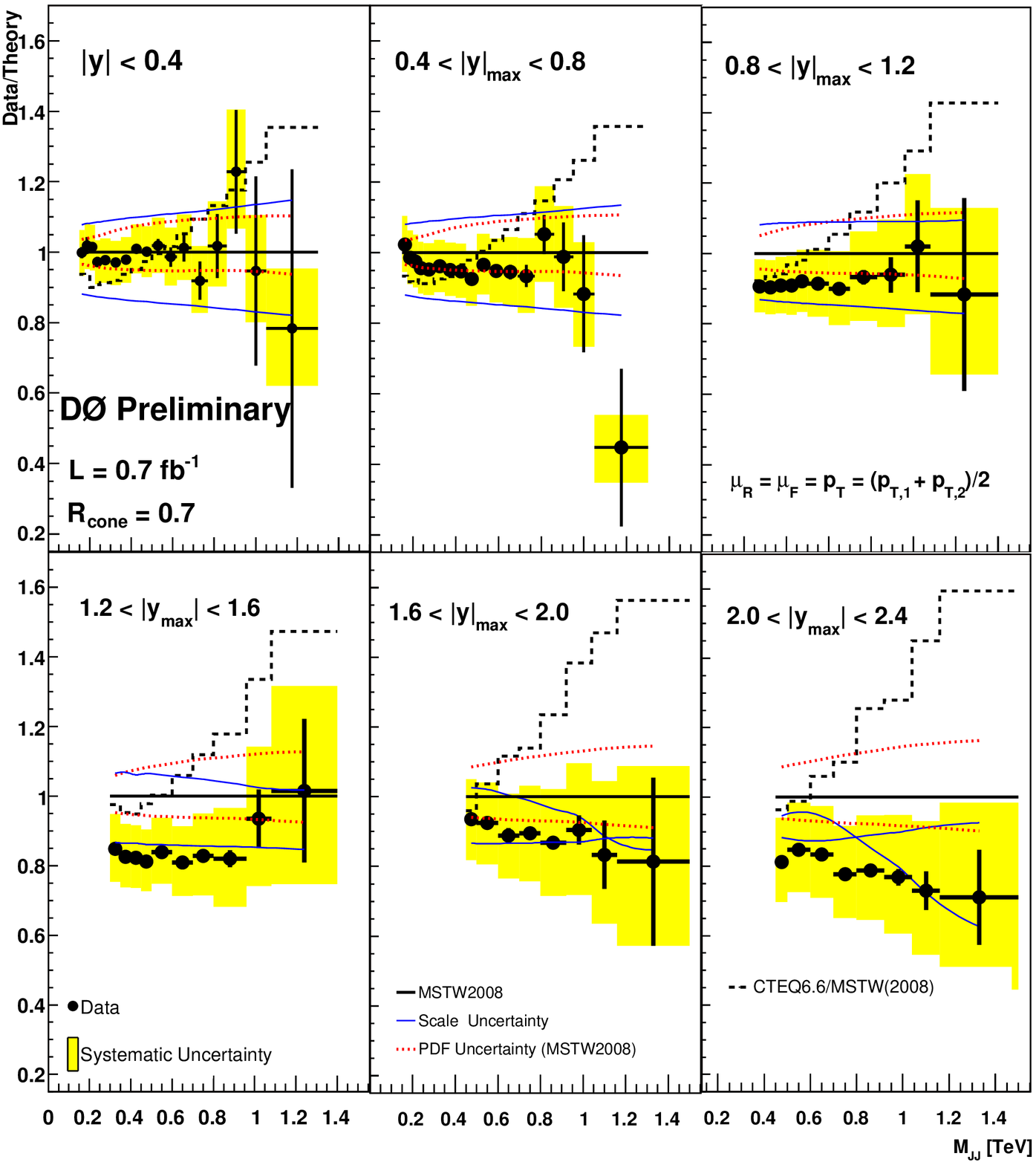}
\caption{{\em (left)} D{\O}'s dijet mass measurement for six rapidity bins~\cite{d0_dijet_mass} compared to NLO using MSTW2008 PDFs.
{\em (right)} Ratio of data/theory.  The experimental systematic uncertainties are very small.  The dashed line shows the effect of using the CTEQ6.6M PDFs.\label{Fig-D0-dijets}}
\end{center}
\end{figure}

The dijet mass spectrum measured by the \D0 collaboration in six rapidity bins~\cite{d0_dijet_mass} is shown in 
Figure~\ref{Fig-D0-dijets}. The rapidity bin is labeled by the higher of the two jet rapidities.
The data are compared to NLO pQCD predictions computed by FastNLO~\cite{FastNLO} using the 
MSTW2008 NLO PDFs~\cite{MSTW} with scale $\mu_R=\mu_F=(p_{T,1}+p_{T,2})/2$ and are corrected for non-perturbative effects
determined using the \pythia event generator. 
The data and QCD predictions are in reasonable agreement for the $|y|<0.4$ region which is not surprising as the 
MSTW2008 global fit includes the inclusive jet data (see Section~\ref{sec-pdfs}). For
higher $|y|$ bins, the data are below the theoretical predictions but within $1\,\sigma$ of
the total experimental systematic uncertainty. The \D0 collaboration is 
searching for new particles using these data.

\subsection{Dijet angular distributions}
The angle between the initial and final state partons in the center of momentum frame
is sensitive to the spin of the exchanged or the intermediate
particle and thus can be used to search for
physics beyond the Standard Model. At hadron colliders, dijet production is 
dominated by the $t$ channel exchange of a gluon, a massless vector boson, and the angular distribution has the familiar Rutherford 
scattering form 
\begin{eqnarray}
\frac{d\hat{\sigma}}{d\cos\theta^*} \sim \frac{1}{(1-cos\theta^*)^2} = \frac{1}{\sin^4(\theta^*/2)}
\end{eqnarray}
where
$\theta^*$ is the angle between the jet and the beam direction in the dijet center of momentum frame. 
The angular distribution of the new particles proposed in many new physics scenarios is relatively flat
in $\cos\theta^*$.  For example, the angular distribution of spin 1 particles (\Wprime, \Zprime, Axigluon, coloron) decaying
in fermions is $d\sigma/d\cos\theta^* \sim  1+cos^2\theta^*$.
Theories in which
quarks are composite particles but with the compositeness scale much higher than the
available energy, can be parametrized by an effective Lagrangian of the type~\cite{compositeness,Eichten-96,Lane-96},
\begin{eqnarray*}
{\cal L} =\eta \frac{g^2}{4\Lambda^2} (\bar{q}_i\gamma^\mu q_i) (\bar{q_j}\gamma_\mu q_j) \hspace{5ex} i= L,R, \;j=L,R
\end{eqnarray*}
where $\Lambda$ is a parameter in the theory which controls the characteristic energy of the new interactions.
The parameter $\eta$ is $\pm 1$ and determines the sign of interference between new interactions and the SM interactions.
The main effect of substructure is to increase the proportion of centrally produced jets, which
can be observed in the jet angular distributions~\cite{Eichten-96}. 

The ADD LED models~\cite{LED-ADD,LED-Atwood}, proposed to solve the hierarchy problem, i.e. the difference between 
electroweak scale ($\sim 100$ GeV) and
the Plank scale  $M_{Pl}$ ($\sim 10^{19}$ GeV), assume the existence of extra spatial dimensions
in which gravity is allowed to propagate. As a consequence, gravity appears weak in the three conventional spatial dimensions. 
The Planck scale, the number of extra dimension $n$, their size $R$ and an effective Planck scale $M_S$ are 
related by $M_{Pl} =  M_S R^{n}$. Experimentally, $M_S$ can be measured for different values of $n$.
The Kaluza-Klein excitations of the graviton
can be exchanged between partons and thus contribute to jet production,  resulting in jets which are central. 
There are two different formalisms to describe LED models, GRW~\cite{LED-GRW} and HLZ~\cite{LED-HLZ}. 
In the HLZ formalism, the sub-leading dependence on the 
number $n$ of extra dimensions is also included. 

In some models~\cite{TevED-Dienes,TevED-Pomarol,TevED-Cheung}, extra dimensions are assumed to  exist at the \TevInv
distance scale. In these models, Kaluza-Klein excitations of SM bosons modify various production cross sections.
In these models, gluons can propagate through the extra dimensions, which changes the jet cross section.
The strength of the interaction is given by the model parameter, the compactification scale, $M_C$.

To search for new physics, instead of studying the $\cos\theta^*$ distribution directly, 
it is convenient to use the \chidijet distribution which removes the Rutherford
singularity: $\chidijet$ is defined as ${\rm exp}(|y_1-y_2|)$ where $y_1$ and $y_2$ are 
rapidities~\cite{kinematics} of the two highest \pT jets in an event. 
For $2\rightarrow 2$ scattering of massless partons, the variable $\chi_{dijet}$ is related to the partonic
center-of-momentum frame polar angle 
$\theta^*$ by $\chi_{dijet}= (1+\cos\theta^*)/(1-\cos\theta^*)$.

\begin{figure}[htbp]
\begin{center}
\includegraphics[width=0.450\hsize]{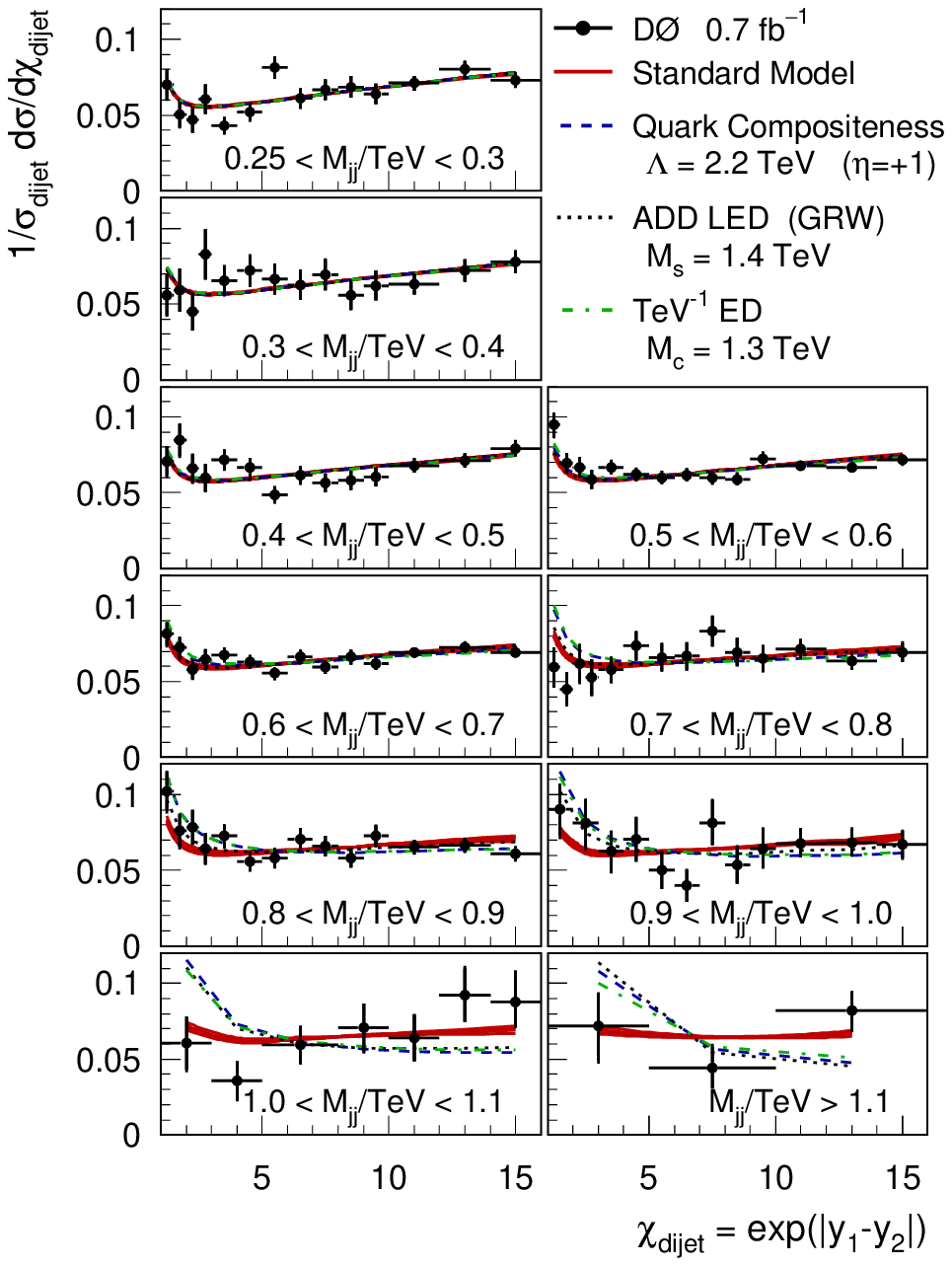}
\caption{D{\O}'s measurement of the dijet $\chi$ for ten dijet mass bins~\cite{d0_dijet_chi}.  The solid line is the Standard Model prediction at NLO.  The dashed lines show the predictions for various new physics models.\label{Fig-D0-chi}}
\end{center}
\end{figure}

The CDF collaboration studied the dijet angular distributions using 106\pbinv of data from Run~I~\cite{cdf_dijet_chi}.
The data excludes at 95\% CL a model of quark substructure in which only up and down quarks are composite and 
the contact interaction scale is $\Lambda_{ud}^{+}\le 1.6$ TeV or $\Lambda_{ud}^{-}\le 1.4$ TeV where the subscript refers to
the flavor of quarks assumed to be composite and the superscript $\pm$ refers to the sign of the interference.
For a model in which all quarks are 
composite, the excluded regions are $\Lambda^{+}\le 1.8$ TeV and $\Lambda^{-}\le 1.6$ TeV.
In Run~II, the \D0 collaboration
measured the \chidijet distribution in the $\chidijet\le 16$ range in 10 dijet mass $m_{jj}$ bins covering the $0.25<m_{jj}<1.1$ \TeVCsqr 
range using up to 0.7 \fbinv of data collected during $2004-2005$~\cite{d0_dijet_chi}. The boost of the two-jet system is required to be
 $y_{\rm boost}\equiv 0.5\times |y_1+y_2| \le 1$.
This requirement, combined with the dijet mass cut and the range of the \chidijet distribution, restricts the highest allowable 
rapidity to $|y_{1,2}|< 2.4$ where the \D0 detector performance is well understood.
The measured distributions are corrected for detector effects using events generated with {\pythia}
v6.419~\cite{pythia} with tune QW~\cite{QW-tune} and the MSTW2008LO parton distributions functions. This procedure corrects for the migration
between dijet mass bins, as well as the shape of the \chidijet distributions in each mass bin. 
The corrected normalized differential cross section 
distributions
$(1/\sigma_{dijet}\cdot d\sigma/d\chidijet)$ at the particle level are shown 
in Figure~\ref{Fig-D0-chi} for 10 dijet mass bins. The NLO pQCD predictions are computed using FastNLO~\cite{FastNLO} based on NLOJet++~\cite{NLOJet++}.
These parton level predictions are corrected for the hadronization and underlying event contributions, which are evaluated
using {\pythia}. The theoretical uncertainties on the SM \chidijet distributions arising from the uncertainty on the PDFs and
the uncertainty on renormalization and factorization scales are less than 2\% and 5\% respectively. 
These data are in good agreement with the SM predictions and thus are used to set exclusion limits in the parameter space
of quark compositeness, ADD LED and \TevInv models.
Calculations for all these models are available only at leading order, while pQCD
calculations can be performed at next to leading order. For this analysis, the expected distributions for each new model 
are calculated at LO and then scaled by $k$-factors ($k=\sigma_{\rm NLO}/\sigma_{\rm LO}$) determined from pQCD calculations.
The $k$-factors vary from 1.25 to 1.5. All these models predict a higher rate as $\chidijet\rightarrow 1$ and as $m_{jj}$ increases.
However, the magnitude of the excess is different for different models.
A Bayesian procedure~\cite{D0-Bayesian} is used to obtain 95\% C.L. limits on the mass scale parameters
$\Lambda$, $M_C$ and $M_S$ in the above models.
The results in which the prior is chosen to be flat 
in the model cross section are given in Table~\ref{Table-chi-limits}. 
Other choices give similar but slightly higher limits~\cite{d0_dijet_chi}.

The limits on $M_C$ obtained in this analysis are the first direct search for \TevInv extra dimensions at a 
particle collider, though inferior to indirect limits from precision electroweak measurements~\cite{TevED-Cheung}.
The limits on $M_S$ in different formalisms of ADD LED are on average slightly higher than the
recent \D0 results obtained using dielectron and diphoton data~\cite{D0-diphoton}. The quark compositeness limits are the most stringent limits
to date.

\begin{table}
\begin{center}
\caption{\label{Table-chi-limits}Expected and observed 95\% C.L. limits on various new physics models.}
\begin{tabular}{lcc}
\hline
\hline
Model (parameter)   & Expected (TeV)   & Observed  (TeV) \\
\hline
Quark Compositeness ($\Lambda$)  &   &  \\
$\eta = +1$         & 2.76 & 2.84 \\
$\eta = -1$         & 2.75 & 2.82 \\
\hline
\TevInv ED ($M_C$)  & 1.60 & 1.55 \\
\hline
ADD  LED ($M_S$)    &   & \\
GRW                 & 1.47 & 1.59 \\
HLZ $n=3$           & 1.75 & 1.89 \\
HLZ $n=4$           & 1.47 & 1.59 \\
HLZ $n=5$           & 1.33 & 1.43 \\
HLZ $n=6$           & 1.24 & 1.34 \\
HLZ $n=7$           & 1.17 & 1.26 \\
\hline
\hline
\end{tabular}
\end{center}
\end{table}


\section{Conclusions}
Since the start of the Run~II, there has been a significant increase in the experimental data used in the
jet analyses at the Tevatron. Due to the higher production cross section and the increase in available integrated
luminosity, the inclusive jet cross section measurement has been extended to transverse momenta of  600 \GeVC.
The experimental uncertainty is still dominated by the uncertainty on the jet energy scale which is
+31/-26\%  in the $\pT=457-527$ \GeVC bin in $0.1<|y|<0.7$ region for the CDF measurements and
+16.0/-15.5\% in the $\pT=490-540$ \GeVC bin in $|\y|<0.5$ region for \D0 measurements.
Both collaborations used a modern
jet finding algorithm, the midpoint algorithm, which is infra-red and collinear safe at next-to-leading order in pQCD for 
measurements of the inclusive jet cross section. In addition, the treatment of the underlying event energy and hadronization
effects has improved over the techniques used in Run~I. The experimental uncertainties are lower than the theoretical
uncertainties, which are dominated by the uncertainties on the parton distribution functions. These data have been used in
global fits to determine the parton distributions and have decreased the uncertainty on the gluon distribution
for the $x\ge 0.3$ region. The new gluon distribution is slightly lower than that determined by Run~I jet cross section
measurements. This increase in the accuracy of gluon distributions will make the prediction of various
processes  more precise.

These jet data have been successfully used to search for physics beyond the Standard Model using jet kinematic distributions.
The dijet mass spectrum has been used to expand the exclusion regions in parameters of excited quarks in quark 
compositeness models, $E_6$ diquarks, axigluons and heavy vector 
bosons $W$ and $Z$ bosons, and the techi-$\rho$ in color octet models. The 95\% C.L. lower mass limits range from 630 \GeVCsqr for colorons to
1.25 \TeVCsqr for axigluons.
The dijet angular distribution has been used to extend the limits
on the quark compositeness mass scale, ADD large extra dimensions and the \TevInv extra dimensions.
The 95\% C.L. lower limit on the compositeness  mass scale is  2.8 TeV. 
The 95\% C.L. lower limit for compactification mass scale in the \TevInv model is 1.5 TeV.
The limits on the ADD large extra dimensions range from 1.9 TeV to 1.3 TeV depending on the number 
of extra dimensions in the HLZ formalism. The 95\% C.L. in the GRW formalism is 1.6 TeV. In most cases, these are
the best limits to date. 

We thank Andrew Beretvas, Kenichi Hatakeyama and Marek Zielinski for reading the manuscript and for insightful comments.


\begin{thebibliography}{99}
\bibitem{kinematics} The important kinematic variables are the rapidity $y = \frac{1}{2} \ln \frac{E + p_z}{E - p_z}$, the azimuthal angle $\phi = \tan^{-1} \frac{p_y}{p_x}$ and transverse momentum $p_T = \sqrt{p_x^2 +p_y^2}$ where a particle or jet has a 4-momentum $(E, p_x, p_y, p_z)$ and the positive $z$ is defined by the proton beam's direction, positive $y$ is up and the $x$ direction is that required to make a right-handed coordinate system.  In addition, the pseudorapidity $\eta = - \ln [\tan (\theta/2)]$ is sometimes used, where $\theta$ is the polar angle, measured with respect to the proton beam axis.  The boost rapidity is defined as $y_{\rm boost} = 0.5|y_1+y_2|$.
\end{thebibliography}
\end{document}